\documentclass[lettersize,journal]{IEEEtran}
\usepackage{amsmath,amsfonts}
\usepackage{amssymb}
\usepackage{algorithmic}
\usepackage{algorithm}
\usepackage{array}
\usepackage[caption=false,font=normalsize,labelfont=sf,textfont=sf]{subfig}
\usepackage{textcomp}
\usepackage{stfloats}
\usepackage{url}
\usepackage{verbatim}
\usepackage{graphicx}
\usepackage{cite}
\usepackage{color} 
\usepackage{makecell}

\allowdisplaybreaks[4]

\begin{document}

\title{Optimal Beamforming for MIMO DFRC Systems with Transmit Covariance Constraints}

    \author{Chenhao Yang, Xin Wang,~\IEEEmembership{Fellow,~IEEE}, Wei Ni,~\IEEEmembership{Senior Member,~IEEE}, and Yi Jiang,~\IEEEmembership{Member,~IEEE}
\thanks{C. Yang, X. Wang, and Y. Jiang are with the Department of Communication Science and Engineering, Fudan University, Shanghai 200433, China (email:xwang11@fudan.edu.cn). }
\thanks{W. Ni is with the Data61, Commonwealth Scientific and Industrial Research Organization (CSIRO), Sydney, Marsfield, NSW 2122, Australia (e-mail:wei.ni@data61.csiro.au).} 
}

\markboth{}%
{Shell \MakeLowercase{\textit{et al.}}: A Sample Article Using IEEEtran.cls for IEEE Journals}


\maketitle

\begin{abstract}
This paper optimizes the beamforming design of a downlink multiple-input multiple-output (MIMO) dual-function radar-communication (DFRC) system to maximize the weighted communication sum-rate under a prescribed transmit covariance constraint for radar performance guarantee. In the single-user case, we show that the transmit covariance constraint implies the existence of inherent orthogonality among the transmit beamforming vectors in use. Then, leveraging Cauchy's interlace theorem, we derive the globally optimal 
transmit and receive beamforming solution in closed form. In the multi-user case, we exploit the connection between the weighted sum-rate and weighted minimum mean squared error (MMSE) to reformulate the intended problem, and develop a block-coordinate-descent (BCD) algorithm
to iteratively compute the transmit beamforming and receive beamforming solutions. Under this approach, we reveal that the optimal receive beamforming is given by the classic MMSE one and the optimal transmit beamforming design amounts to solving an orthogonal Procrustes problem, thereby allowing for closed-form solutions to subproblems in each BCD step and fast convergence of the proposed algorithm to a high-quality (near-optimal) overall beamforming design. Numerical results demonstrate the superiority of our approach to the existing methods, with at least 40\% higher sum-rate under a multi-user MIMO setting in the high signal-to-noise regime.
\end{abstract}


\section{Introduction}
\IEEEPARstart{T}{o}  meet the rapidly growing demand for high-speed mobile data services, conventional radar spectrum bands, such as S-band (2 to 4 GHz) and C-band (4 to 8 GHz), are nowadays resumed for communication applications. To this end, integrated sensing and communications (ISAC) has drawn growing attention in the context of the beyond five-generation (B5G) and sixth-generation (6G) communication system designs~\cite{1,2,mag3}.

Earlier approaches to ISAC investigated the co-existence of radar and communication systems, where radar and communication systems operating on different devices exchange side information for cooperation. In \cite{3}, radar beamformers were designed to project radar signals onto the null space of the interference channels between the radar and the base station (BS), thereby zero-forcing the interference imposed on the communication
link. The authors of \cite{5,5a} optimized the radar beamformers and communication covariance matrix jointly, to maximize the radar’s SINR under communication performance and power budget constraints. Under a similar setting, a robust beamforming design with imperfect channel state information (CSI) was investigated in \cite{5b}. Moreover, full-duplex multi-user communications were considered in~\cite{fulldup} to improve spectrum efficiency. Although the co-existence of radar and communication appears to be an effective way to support spectrum sharing in the same band, these separate radar and communication system designs would incur extra hardware and energy costs, as well as increased system complexity~\cite{6}.

As a more efficient design, a recent approach to spectrum sharing between communication and radar sensing in the same band is to develop dual-function radar-communication (DFRC) systems, where the same set of signals is used to serve both purposes of sensing and communications on the same hardware platform. With relatively low system complexity and energy cost~\cite{6}, DFRC systems can be installed, e.g., in autonomous vehicles to sense environments and communicate via vehicle-to-everything (V2X) links~\cite{mv} and in unmanned aerial vehicle (UAV) networks to incorporate identity production, mapping, management, and authentication \cite{uav}.

A majority of the existing literature on DFRC systems has aimed to optimize radar performance under communication performance constraints, leading to a communication-centric design. In~\cite{6} and~\cite{7}, semidefinite relaxation (SDR)-based approaches were developed to obtain beamforming designs that minimize radar beampattern matching errors, subject to the users’ communication quality-of-service (QoS) constraints. The mean square error (MSE) for estimating the target response matrix~\cite{liufanconfer} and the Cramér–Rao bound (CRB) for estimating the direction-of-arrival (DoA)~\cite{8,18} were minimized under communication performance constraints. Furthermore,  the authors of \cite{10} developed joint transmit and receive beamforming designs to maximize the signal-to-interference-plus-noise ratio (SINR) at the radar receivers. In general, these communication-centric designs~\cite{6,7,8,qichenhao,liufanconfer,10,18} could only offer best-effort radar sensing performances, and the sensing performance loss was thus inevitable. To better preserve radar performance, communication performance was optimized under radar beampattern mismatch errors constraint, e.g., in reconfigurable intelligent surface (RIS)-aided systems~\cite{ris} and secure transmission systems~\cite{secure1}. However, the above-mentioned designs, i.e.,~\cite{ris,secure1}, could still somewhat impair the sensing performance.


 By contrast, radar-centric DFRC system designs incorporate communication functions into pre-designed radar systems to guarantee radar performance. To this end, some prior works have attempted to embed communication information into radar signals to achieve dual functionality without degrading
radar performance~\cite{yimin,coexmag,indexmodu}. 
For instance,  in \cite{yimin}, the sidelobe levels of radar pulses were modulated using amplitude shift keying (ASK), where different power levels represent different communication symbols. In~\cite{indexmodu}, different sets of frequencies were allocated to different antennas by antenna permutation to convey communication information. However, spatial diversity was not utilized in these studies.

Since the sensing performance relies heavily on the transmit covariance of waveforms~\cite{2008tae,lijian}, a few recent works optimized the communication performance under a prescribed transmit covariance constraint for a stringent radar performance guarantee. The authors of \cite{11} optimized the instantaneous transmit waveform under a prescribed transmit covariance constraint for multi-user multiple-input single-output (MISO) systems. Under the same constraint, optimal beamforming designs were put forth in \cite{12}, where dirty paper coding (DPC) was leveraged to improve communication performance when each communication user only receives a single data stream in the downlink. Nevertheless, the DPC scheme could incur significant implementation and computing complexity~\cite{SDR}, and the approaches in~\cite{11,12} cannot be readily extended to more general multiple-input multiple-output (MIMO) DFRC systems where transmissions of multiple data streams are allowed for each communication user.


This paper considers a  downlink MIMO system where a  BS simultaneously probes multiple targets while transmitting information to multiple communication users in parallel. For such a DFRC system, we aim to develop the optimal beamforming design that maximizes the weighted sum-rate for communicating users under the prescribed transmit covariance constraint to guarantee radar performance.

The contributions of the paper are summarized as follows.
\begin{itemize}
    \item 
    We propose novel approaches to (near-)optimal beamforming designs for DFRC systems that can effectively leverage the multi-stream MIMO transmissions (i.e., MIMO multiplexing gain) to achieve maximum communication rates, while strictly preserving the radar sensing performance (enforced by the transmit covariance constraint).
    \item 
    In the single-user MIMO DFRC case, we show that there exists inherent orthogonality among the transmit beamforming vectors due to the transmit covariance constraint, and then apply a corollary of Cauchy’s interlace theorem to rigorously prove that the globally optimal transmit and receive beamformers can be analytically obtained in closed form.
    \item
    In the multi-user MIMO DFRC case, we exploit the connection between weighted sum-rate and weighted minimum mean squared error (MMSE) to reformulate the intended problem and develop an efficient block-coordinate-descent (BCD) type method to iteratively refine the transmit and receive beamforming solutions.
    \item
    With the proposed BCD approach, we reveal that the optimal receive beamforming is the classic MMSE one and the optimal transmit beamforming design amounts to solving an orthogonal Procrustes problem. Closed-form solutions can then be derived for the subproblems in each BCD step, ensuring fast convergence of our algorithm to a high-quality (near-optimal) overall beamforming design.

    
\end{itemize}

\begin{table}[]
\caption{Notation and Definitions}
\centering
\renewcommand{\arraystretch}{1.3}
\label{tab:my-table}
\small
\begin{tabular}{l|llll}
\cline{1-2}
\textbf{Notation} & \textbf{Definition}                   &  &  &  \\ \cline{1-2}
$N_{\rm tx}$      & Number of transmit antennas at the DFRC BS; &  &  &  \\
$K$               & Number of users;  
&  &  &  \\
$N_{\rm rx}$      & Number of receive antennas at each user;  
&  &  &  \\
$d$      & Number of transmit symbols sent to each user;  
&  &  &  \\
${\mathbf{x}}(t)$      & Transmit signal vector at the DFRC BS;  
&  &  &  \\
${{\mathbf{F}}_k}$      & Transmit beamformer of the $k$-th user at the  
&  &  &  \\
                        & DFRC BS; 
&  &  &  \\
${{\mathbf{F}}_{\rm c}}$      & Communication transmit beamformer at the DFRC  
&  &  &  \\
                             &  BS, ${{\mathbf{F}}_{\rm c}} = [{{\mathbf{F}}_1},{{\mathbf{F}}_2},\cdots,{{\mathbf{F}}_K}]$; 
&  &  &  \\
${{\mathbf{F}}_{\rm r}}$      & Radar transmit beamformer at the DFRC BS; 
&  &  &  \\
$\mathbf{F}$      & Overall transmit beamformer at the DFRC BS, 
&  &  &  \\
                  & ${\mathbf{F}} = [{{\mathbf{F}}_{\rm c}},{{\mathbf{F}}_{\rm r}}]$; 
&  &  &  \\
$\mathbf{\tilde{F}}$      &Auxiliary variables representing equivalent transmit  
&  &  &  \\
                          & beamforming matrices;
&  &  &  \\
$\mathbf{G}_k$      & Receive beamformer at the $k$-th user; 
&  &  &  \\
$\mathbf{G}$        & Collections of  all users' receive beamforming  
&  &  &  \\
                    &   matrices;
&  &  &  \\
$\mathbf{H}_k$      & Channel matrix from the BS to the $k$-th user;
&  &  &  \\
$\mathbf{\tilde{H}}_k$      &Auxiliary variables representing  equivalent 
&  &  &  \\
                            &  channel matrices;
&  &  &  \\
$\mathbf{W}_k$        &  Weight matrix of the $k$-th user;
&  &  &  \\
$\mathbf{W}$        &  Collections of  all users' weight matrices;
&  &  &  \\
$\mathbf{E}_k$        & MSE matrix  of the $k$-th user;
&  &  &  \\
$\theta$        & Angular directions of targets;
&  &  &  \\
${{\mathbf{R}}_{\rm{des}}}$      & Prescribed transmit covariance matrix for 
&  &  &  \\
                                 & guaranteed  sensing performance; 
&  &  &  \\
${\mathbf{L}}$      & Lower triangular matrix achieved by Cholesky  
&  &  &  \\ 
                    &   decomposition ${{\mathbf{R}}_{\rm{des}}} = {\mathbf{L}}{{\mathbf{L}}^H}$.
&  &  &  \\ 
\cline{1-2}
\end{tabular}%
\end{table}

Extensive simulations demonstrate that the globally optimal beamforming design is achieved in the single-user case. It is also shown that the proposed BCD algorithm converges fast to an efficient beamforming design solution that is at least 40\% better than benchmark schemes under high signal-to-noise ratio (SNR) settings in the multi-user case. The performance trade-offs between sensing and communication are also studied.

The rest of this paper is organized as follows. Section II describes the system model. Section III derives the globally optimal beamforming design in the single-user case. Section IV develops an efficient BCD-type algorithm to obtain a high-quality beamforming design in the multi-user case. Simulation results are provided in Section V, followed by the conclusions in Section VI.

{\em Notation:} Bold-face lower- and upper-cases indicate vectors and matrices, respectively; ${( \cdot )^T}$, ${( \cdot )^H}$, $( \cdot )^C$, ${\rm{Tr(\cdot)}}$, ${\rm{det(\cdot)}}$, $|| \cdot |{|_F}$, and ${\cal R( \cdot )}$ stand for transpose, conjugate transpose, conjugation, trace,  determinant, Frobenius norm, and range operator, respectively; ${\lambda _m}( \cdot )$ represents the $m$-th smallest eigenvalue of a matrix; ${\mathbf{I}_n}$ denotes the $n \times n$ identity matrix;  ${\mathbb{C}^{m \times n}}$ denotes the set of $m \times n$ complex matrices; $\operatorname{Re} \{  \cdot \}$ takes the real part of a complex value; $\mathbb{E}\{  \cdot \}$ denotes ensemble expectation.

\section{System Model}
We consider a  MIMO DFRC system, where a BS simultaneously probes $J$ far-field targets and transmits signals to $K$ (communication) users in the downlink. The BS is equipped with $N_{\rm tx}$ transmit antennas. Each user is equipped with $N_{\rm rx}$ ($N_{\rm rx} \leqslant N_{\rm tx}$) receive antennas.  The channel matrix from the BS to user $k$ is denoted by ${{\mathbf{H}}_k} \in {\mathbb{C}^{N_{\rm rx} \times N_{\rm tx}}}$.

In terms of communications, let ${{\mathbf{F}}_k} \in {\mathbb{C}^{{N_{\rm tx}} \times d}}$ denote the beamforming matrix for (multi-stream) communication symbol vector ${{\mathbf{c}}_k}(t) \in {\mathbb{C}^{d \times 1}}$ at time $t$ from BS to user $k$. (Clearly, the classic information-theoretic result on MIMO multiplexing dictates that we should have $d \leq \min\{N_{\rm tx}, N_{\rm rx}\}$.) Without loss of generality, we assume that the transmit symbols of different users are generated independently and $\mathbb{E}\{ {{\mathbf{c}}_k}(t){{\mathbf{c}}_k}^H(t)\}  = {{\mathbf{I}}_d}$. The transmit symbol vector ${\mathbf{c}}(t)= [ \mathbf{c}_1^T(t), \cdots, \mathbf{c}_K^T(t)]^T$ containing $D = d \times K$  data streams is precoded linearly by a beamforming matrix ${{\mathbf{F}}_{\rm c}} = [{{\mathbf{F}}_1},{{\mathbf{F}}_2},\cdots,{{\mathbf{F}}_K}] \in {\mathbb{C}^{{N_{\rm tx}} \times D}}$.
On the receiver side, let $\mathbf{G}_k \in \mathbb{C}^{N_{\rm rx} \times d}$ denote the receive beamformer of user $k$; and $\mathbf{G}= [\mathbf{G}_1, \mathbf{G}_2,\cdots, \mathbf{G}_K]$ collects the receive beamforming matrices of all $K$ users.

 In terms of sensing, a radar signal vector ${\mathbf{r}}(t) \in {\mathbb{C}^{({N_{\rm tx}} - D) \times 1}}$ consisting of  $(N_{\rm tx}-D)$ independently and pseudo-randomly generated symbols\cite{7}, is transmitted along with $\boldsymbol{c}(t)$. Assume that  ${\mathbb{E}\{ {\mathbf{r}}(t){{\mathbf{r}}^H}(t)\}  = {{\mathbf{I}}_{N_{\rm tx} - D}}} $, and radar signals are uncorrelated with the communication symbols, i.e., $\mathbb{E}\{ {\mathbf{r}}(t){{\mathbf{c}}^H}(t)\}  = {{\mathbf{0}}_{(N_{\rm tx} - D) \times D}}$. The radar symbols are precoded by a beamforming matrix ${{\mathbf{F}}_{\rm r}} \in {\mathbb{C}^{N_{\rm tx} \times (N_{\rm tx} - D)}}$. 
 
With the $N_{\rm tx} \times N_{\rm tx}$ beamforming matrix  ${\mathbf{F}} = [{{\mathbf{F}}_{\rm c}},{{\mathbf{F}}_{\rm r}}]$,  the transmit signal vector  ${\mathbf{x}}(t) \in {\mathbb{C}^{{N_{\rm tx}} \times 1}}$ is then given by 
\begin{equation} \label{eq1}
{\mathbf{x}}(t) = {\mathbf{F}}\left[ {\begin{array}{*{20}{c}}
  {{\mathbf{c}}(t)} \\ 
  {{\mathbf{r}}(t)} 
\end{array}} \right] = {{\mathbf{F}}_{\rm c}}{\mathbf{c}}(t) + {{\mathbf{F}}_{\rm r}}{\mathbf{r}}(t),{\text{ }}t = 0,1, \cdots .
\end{equation}

\subsection{Radar Performance Guarantee}
 For a MIMO radar, the sensing performance depends highly on the transmit beampattern. Based on the signal model (\ref{eq1}),  the transmit beampattern is determined by the covariance of transmit signals, i.e., 
\begin{equation} 
{{\mathbf{R}}_x} = \mathbb{E}\{ {\mathbf{x}}(t){{\mathbf{x}}^H}(t)\}  = {\mathbf{F}}{{\mathbf{F}}^H} = {{\mathbf{F}}_{\rm c}}{\mathbf{F}}_{\rm c}^H + {{\mathbf{F}}_{\rm r}}{\mathbf{F}}_{\rm r}^H.
\end{equation}
Achieving the desired beampattern then amounts to a transmit covariance constraint for the beamforming matrix.
In particular, depending on the specific sensing requirement, a desired transmit covariance matrix, denoted by ${{\mathbf{R}}_{\rm{des}}}$, can be obtained in advance \cite{lijian}. To ensure an acceptable sensing performance, we would then require the transmit covariance to match the prescribed ${{\mathbf{R}}_{\rm{des}}}$; i.e., 
\begin{equation} \label{eq2}
{\mathbf{F}}{{\mathbf{F}}^H} = {{\mathbf{R}}_{\rm{des}}}.
\end{equation}

It is worth noting that the transmit covariance constraint (\ref{eq2}) here indeed also implicitly provides the power constraint for the transmit beamforming matrix $\mathbf{F}$.  Given $\mathbf{F}$, the total transmit power is clearly given by $P = {\text{Tr}}({\mathbf{F}}{{\mathbf{F}}^H})$. To explicitly present the power constraint as in the traditional beamforming design problem, we may define a normalized $\tilde{\mathbf{R}}_{\text{des}}$ with $\text{Tr}(\tilde{\mathbf{R}}_{\text{des}})=1$, and rephrase the transmit covariance constraint as $\frac{{\mathbf{F}{\mathbf{F}}^H}}{\text{Tr}({\mathbf{F}{\mathbf{F}}^H})}=\tilde{\mathbf{R}}_{\text{des}}$, along with the power constraint
\begin{equation}
{\text{Tr}}({\mathbf{F}{\mathbf{F}}^H}) \leq P_{\text{max}}
\end{equation}
where $P_{\text{max}}$ is the power budget at the BS. As radars always transmit probing waveforms with their maximal available power\cite{lijian} and our objective function, i.e., the weighted sum of user rates defined in the sequel, is monotonically non-decreasing with the total transmit power, the normalized transmit covariance constraint and the power constraint can be equivalently combined into the one in (\ref{eq2}), with ${\mathbf{R}}_{\text{des}} \equiv {P_{\text{max}}}{\tilde{\mathbf{R}}_{\text{des}}}$. For conciseness, hereinafter, we simply use the compact form (\ref{eq2}) in our problem formulation.

\subsection{Communication Performance Metric}
For the downlink communications of the DFRC system, the received signal of user $k$ is  the mixture of its own signal, the interference (which are the signals intended for the other users), the radar signal, and the receiver noise, as given by
\begin{equation} 
 {{\mathbf{y}}_k}(t) = {{\mathbf{H}}_k}{{\mathbf{F}}_k}{{\mathbf{c}}_k}(t) + {{\mathbf{H}}_k}\sum\limits_{i \ne k} {{{\mathbf{F}}_i}{{\mathbf{c}}_i}(t)}  + {{\mathbf{H}}_k}{{\mathbf{F}}_{\rm r}}{\mathbf{r}}(t) + {{\mathbf{n}}_k}(t), 
\end{equation}
where ${{\mathbf{n}}_k}(t) \in {\mathbb{C}^{N_{\rm rx} }}$ denotes the additive white Gaussian noise (AWGN) with zero mean and covariance matrix $\sigma ^2{\mathbf{I}_{N_{\rm rx}}}$. The noise is independent of communication and radar signals. 

With a linear receive beamforming matrix ${{\mathbf{G}}_k} \in {\mathbb{C}^{{N_{\rm rx}} \times d}}$, the estimated signal at user $k$ is given by
\begin{equation} 
{{\mathbf{\hat c}}_k}(t) = {\mathbf{G}}_k^H{{\mathbf{y}}_k}(t).
\end{equation}
We use the weighted sum of user rates as the communication performance metric. Specifically, the weighted sum-rate of the considered MIMO DFRC system is expressed as
\begin{equation}  \label{eq3}
C = \sum\limits_{k = 1}^K {{\omega_k}} {C_k},
\end{equation}
where ${{\omega_k}}$ is the weight of user $k$, indicating the priority of the user. Based on the well-known MIMO capacity formula~\cite{14}, the maximum 
achievable rate ${C_k}$ of user $k$ for a given transmit beamforming matrix $\mathbf{F}$ is
\begin{equation} \label{eq3a}
\begin{aligned}
  {C_k} &= \log \det ({{\mathbf{I}}_d} + {\mathbf{F}}_k^H{\mathbf{H}}_k^H 
  ({\sigma ^2} {{\mathbf{I}}_{{N_{\rm rx}}}}    
   + \sum\limits_{i \ne k}^{} {{{\mathbf{H}}_k}{{\mathbf{F}}_i}{\mathbf{F}}_i^H{\mathbf{H}}_k^H}  \\ 
   &+ {{\mathbf{H}}_k}{{\mathbf{F}}_{\rm r}}{\mathbf{F}}_{\rm r}^H{\mathbf{H}}_k^H{)^{ - 1}}{{\mathbf{H}}_k}{{\mathbf{F}}_k}). \\ 
\end{aligned} 
\end{equation}
Note that to achieve the maximum rate in (\ref{eq3a}), the optimal MMSE beamforming matrix $\mathbf{G}_k^{\text{mmse}}$ should be adopted in the receiver. Such a beamforming matrix $\mathbf{G}_k^{\text{mmse}}$ is indeed a function of the given transmit beamforming matrix $\mathbf{F}$; see  (\ref{eq11}) in the sequel. Hence, the expression for the maximum achievable user rate here does not explicitly include the receive beamformer; it can be written merely as a function of $\mathbf{F}$. 

For the MIMO DFRC system under consideration, our goal is then to design the optimal transmit beamformers ${{\mathbf{F}}_{\rm c}}$ and ${{\mathbf{F}}_{\rm r}}$ that maximize the weighted sum-rate (\ref{eq3})  under the transmit covariance constraint (\ref{eq2}).

\section{Optimal Beamforming Design in Single-User Case}
Consider first the single-user case; i.e., there is only $K=1$ user. Then (\ref{eq3a}) reduces to:
\begin{equation}   \label{eq4}
C = \log \det ({{\mathbf{I}}_d} + {\mathbf{F}}_{\rm c}^H{{\mathbf{H}}^H}{({\sigma ^2}{{\mathbf{I}}_{{N_{\rm rx}}}} + {\mathbf{H}}{{\mathbf{F}}_{\rm r}}{\mathbf{F}}_{\rm r}^H{{\mathbf{H}}^H})^{ - 1}}{\mathbf{H}}{{\mathbf{F}}_{\rm c}}),
\end{equation}
where the subscript ``$_k$" is suppressed for conciseness. 

Based on (\ref{eq4}), the problem of interest is formulated as
\begin{subequations} \label{eq5}
\begin{align}
   \mathop {\max }\limits_{\mathbf{F}} {\text{ }} & \log \det ({{\mathbf{I}}_d}{\mathbf{ + F}}_{\rm c}^H{{\mathbf{H}}^H}{({\sigma ^2}{{\mathbf{I}}_{{N_{\rm rx}}}}{\mathbf{ + H}}{{\mathbf{F}}_{\rm r}}{\mathbf{F}}_{\rm r}^H{{\mathbf{H}}^H})^{ - 1}}{\mathbf{H}}{{\mathbf{F}}_{\rm c}})  \\
  {\text{s}}{\text{.t}}{\text{. }}&{\mathbf{F}}{{\mathbf{F}}^H} = {{\mathbf{R}}_{\rm{des}}}.
\end{align}
\end{subequations}
Recall that the total power constraint for $\mathbf{F}$ is actually incorporated into the desired transmit covariance ${{\mathbf{R}}_{{\text{des}}}}$; hence, there is no need for an explicit transmit power constraint as in traditional beamforming design problems, e.g., the one formulated in \cite[Eq.~(4)]{cioffi}. 

It is clear that (\ref{eq5}) is a non-convex problem due to its non-convex objective and quadratic equality constraint. Yet, we next show that the globally optimal $\mathbf{F}^*$ for (\ref{eq5}) can be derived in closed form. To this end, we first perform Cholesky decomposition on ${{\mathbf{R}}_{\rm{des}}}$ to obtain
\begin{equation}  \label{eq6}
{{\mathbf{R}}_{\rm{des}}} = {\mathbf{L}}{{\mathbf{L}}^H},
\end{equation}
where ${\mathbf{L}} \in {\mathbb{C}^{N_{\rm tx} \times N_{\rm tx}}}$ is a lower triangular matrix. 

By substituting (\ref{eq6}) into (\ref{eq5}b), we can rewrite (\ref{eq5}b) as
\begin{equation}  
{{\mathbf{L}}^{ - 1}}{\mathbf{F}}{{\mathbf{F}}^H}{{\mathbf{L}}^{ - H}} = {{\mathbf{I}}_{N_{\rm tx}}}.
\end{equation}
Define ${\mathbf{\tilde F}} = {{\mathbf{L}}^{ - 1}}{\mathbf{F}}$ and ${\mathbf{\tilde H}} = {\mathbf{HL}}$. We can then equivalently rewrite problem (\ref{eq5}) as
\begin{subequations}  \label{eq7}
\begin{align}
  \mathop {\max }\limits_{\mathbf{\tilde F}} &\, \log \det ({{\mathbf{I}}_d}{\mathbf{ + \tilde F}}_{\rm c}^H{{\mathbf{\tilde H}}^H}{({\sigma ^2}{{\mathbf{I}}_{{N_{\rm rx}}}}{\mathbf{ + \tilde H}}{{\mathbf{\tilde F}}_{\rm r}}{\mathbf{\tilde F}}_{\rm r}^H{{\mathbf{\tilde H}}^H})^{ - 1}}{\mathbf{\tilde H}}{{\mathbf{\tilde F}}_{\rm c}})  \\
  {\text{s}}{\text{.t}}{\text{. }}&{\mathbf{\tilde F}}{{\mathbf{\tilde F}}^H} = {{\mathbf{I}}_{N_{\rm tx}}} .
\end{align}
\end{subequations}
From (\ref{eq7}b), the transmit covariance constraint indeed implies inherent orthogonality among the beamforming vectors. This structure facilitates the derivation of the closed-form optimal beamforming solution in the sequel.

To solve (\ref{eq7}), we need a corollary of Cauchy's interlace theorem, as stated below:

\emph{\textbf{Lemma 1} \cite[Corollary 4.3.37]{matrix}\textbf{:} Assume that ${{\mathbf{A}}} \in {\mathbb{C}^{N \times N}}$ is Hermitian and ${{\mathbf{U}}_N} \in {\mathbb{C}^{N \times M}}$ ($M \leqslant N$) has orthonormal columns. Define ${{\mathbf{B}}} = {\mathbf{U}}_N^H{{\mathbf{A}}}{{\mathbf{U}}_N} \in {\mathbb{C}^{M \times M}}$ and arrange the eigenvalues of ${{\mathbf{A}}}$ and ${{\mathbf{B}}}$ in ascending order. Then,
\begin{equation} 
{\lambda _m}({{\mathbf{A}}}) \leqslant {\lambda _m}({{\mathbf{B}}}), {\text{ }} m = 1, \cdots ,M,
\end{equation}
where the equality holds if and only if ${{\mathbf{U}}_N}$ is composed of the eigenvectors associated with  the $M$ smallest eigenvalues of ${{\mathbf{A}}}$.}

Based on Lemma 1, we can then establish the following main result.   

 \emph{\textbf{Theorem 1:} The optimal solution of the problem in (\ref{eq5}) is given by
 \begin{equation}  \label{eq7.1}
 {{\mathbf{F}}^*} = {\mathbf{L}}{\mathbf{V}},
\end{equation}
where ${\mathbf{V}}$ is the right singular matrix of ${\mathbf{\tilde H}}$ obtained by singular value decomposition (SVD), i.e., ${\mathbf{\tilde H}} = {\mathbf{U\Sigma }}{{\mathbf{V}}^H}$.}
\vspace{2 mm}

\emph{Proof:} For the achievable user rate in (\ref{eq7}a), we actually have
\begin{equation} 
\begin{aligned}
  & \log \det ({{\mathbf{I}}_d}{\mathbf{ + \tilde F}}_{\rm c}^H{{{\mathbf{\tilde H}}}^H}{({\sigma ^2}{{\mathbf{I}}_{N_{\rm rx}}}{\mathbf{ + \tilde H}}{{{\mathbf{\tilde F}}}_{\rm r}}{\mathbf{\tilde F}}_{\rm r}^H{{{\mathbf{\tilde H}}}^H})^{ - 1}}{\mathbf{\tilde H}}{{{\mathbf{\tilde F}}}_{\rm c}})   \\
   \mathop  = \limits^{(a)} & \log \det ({{\mathbf{I}}_{N_{\rm rx}}} + {({\sigma ^2}{{\mathbf{I}}_{{N_{{\text{rx}}}}}} + {\mathbf{\tilde H}}{{\mathbf{\tilde F}}_{\text{r}}}{\mathbf{\tilde F}}_{\text{r}}^H{{\mathbf{\tilde H}}^H})^{ - 1}}{\mathbf{\tilde H}}{{\mathbf{\tilde F}}_{\text{c}}}{\mathbf{\tilde F}}_{\text{c}}^H{{\mathbf{\tilde H}}^H}) \\
   = & \log \det ({({\sigma ^2}{{\mathbf{I}}_{N_{\rm rx}}}{\mathbf{ + \tilde H}}{{{\mathbf{\tilde F}}}_{\rm r}}{\mathbf{\tilde F}}_{\rm r}^H{{{\mathbf{\tilde H}}}^H})^{ - 1}} \times  \\
   &({\sigma ^2}{{\mathbf{I}}_{N_{\rm rx}}}{\mathbf{ + \tilde H}}{{{\mathbf{\tilde F}}}_{\rm r}}{\mathbf{\tilde F}}_{\rm r}^H{{{\mathbf{\tilde H}}}^H} + {\mathbf{\tilde H}}{{{\mathbf{\tilde F}}}_{\rm c}}{\mathbf{\tilde F}}_{\rm c}^H{{{\mathbf{\tilde H}}}^H}))  \\ 
    = &\log \det ({({\sigma ^2}{{\mathbf{I}}_{N_{\rm rx}}}{\mathbf{ + \tilde H}}{{{\mathbf{\tilde F}}}_{\rm r}}{\mathbf{\tilde F}}_{\rm r}^H{{{\mathbf{\tilde H}}}^H})^{ - 1}}({\sigma ^2}{{\mathbf{I}}_{N_{\rm rx}}}{\mathbf{ + \tilde H}}{{{\mathbf{\tilde F}}}}{\mathbf{\tilde F}}^H{{{\mathbf{\tilde H}}}^H}))\\
   \mathop  = \limits^{(b)} &\log \det ({({\sigma ^2}{{\mathbf{I}}_{N_{\rm rx}}}{\mathbf{ + \tilde H}}{{{\mathbf{\tilde F}}}_{\rm r}}{\mathbf{\tilde F}}_{\rm r}^H{{{\mathbf{\tilde H}}}^H})^{ - 1}}({\sigma ^2}{{\mathbf{I}}_{N_{\rm rx}}}{\mathbf{ + \tilde H}}{{{\mathbf{\tilde H}}}^H}))\\
   \mathop  = \limits^{(c)} &\log \det ({\sigma ^2}{{\mathbf{I}}_{N_{\rm rx}}}{\mathbf{ + \tilde H}}{{{\mathbf{\tilde H}}}^H}) - \log \det ({\sigma ^2}{{\mathbf{I}}_{N_{\rm rx}}}{\mathbf{ + \tilde H}}{{{\mathbf{\tilde F}}}_{\rm r}}{\mathbf{\tilde F}}_{\rm r}^H{{{\mathbf{\tilde H}}}^H}),
\end{aligned} 
\end{equation}
where $(a)$ is due to $\det ({\mathbf{I}} + {\mathbf{AB}}) = \det ({\mathbf{I}} + {\mathbf{BA}})$, $(b)$ is based on constraint (\ref{eq7}b), and $(c)$ is obtained since $\det ({{\mathbf{B}}^{ - 1}}{\mathbf{A}}) = \det ({\mathbf{A}})/\det ({\mathbf{B}})$. 

As the term $\log \det ({\sigma ^2}{{\mathbf{I}}_{N_{\rm rx}}}{\mathbf{ + \tilde H}}{{{\mathbf{\tilde H}}}^H})$ is independent of $\mathbf{\tilde F}$, solving problem (\ref{eq7}) is equivalent to solving the following problem:
\begin{equation}  \label{eq7.2}
\begin{aligned}
  \mathop {\min }\limits_{{\mathbf{\tilde F}_{\rm r}}} {\text{ }}&\log \det ({\sigma ^2}{{\mathbf{I}}_{N_{\rm rx}}}{\mathbf{ + \tilde H}}{{{\mathbf{\tilde F}}}_{\rm r}}{\mathbf{\tilde F}}_{\rm r}^H{{{\mathbf{\tilde H}}}^H})  \\
  {\text{s}}{\text{.t}}{\text{. }}&{\mathbf{\tilde F}}_{\rm r}^H{{{\mathbf{\tilde F}}}_{\rm r}} = {{\mathbf{I}}_{N_{\rm tx} - d}}.  
\end{aligned}
\end{equation}
Let ${{\mathbf{H}}_e} = {\sigma ^2}{{\mathbf{I}}_{N_{\rm tx}}} + {{\mathbf{\tilde H}}^H}{\mathbf{\tilde H}}$. Then it follows from Lemma 1 that
\begin{equation} 
{\lambda _m}({{\mathbf{H}}_e}) \leqslant {\lambda _m}({\mathbf{\tilde F}}_{\rm r}^H{{\mathbf{H}}_e}{{\mathbf{\tilde F}}_{\rm r}}),{\text{  }}m = 1, \cdots ,{N_{\rm tx}} - d.
\end{equation}
Given that ${{\mathbf{H}}_e}$ is positive definite, we immediately have 
\begin{equation}  \label{eq7.3}
\log (\prod\limits_{m = 1}^{{N_{\rm tx}} - d} {{\lambda _m}({{\mathbf{H}}_e})} ) \leqslant \log \det ({\mathbf{\tilde F}}_{\text{r}}^H{{\mathbf{H}}_e}{{\mathbf{\tilde F}}_{\text{r}}}),
\end{equation}
where the equality is taken if and only if ${{\mathbf{\tilde F}}_{\rm r}}$ consists of the eigenvectors corresponding to  the $(N_{\rm tx}-d)$ smallest eigenvalues of ${{\mathbf{H}}_e}$. With the SVD , i.e., ${\mathbf{\tilde H}} = {\mathbf{U\Sigma }}{{\mathbf{V}}^H}$, we have 
eigenvalue decomposition ${{\mathbf{\tilde H}}^H}{\mathbf{\tilde H}} = {\mathbf{V}}{\mathbf{\Lambda}} {{\mathbf{V}}^H}$, where $\mathbf{\Lambda} = \mathbf{\Sigma}^2$. It then follows that:
\begin{equation} \label{eq8}
{{\mathbf{H}}_e} = {\mathbf{V}}({\sigma ^2}{{\mathbf{I}}_{N_{\rm tx}}} + {\mathbf{\Lambda}} ){{\mathbf{V}}^H}.
\end{equation}
From (\ref{eq8}), it is clear that ${\mathbf{V}}$ is also the eigenvector matrix of ${{\mathbf{H}}_e}$\footnote{In accordance with the convention in linear algebra textbooks, hereinafter we assume that the eigenvalues or singular values are always arranged in descending order in eigenvalue or singular value decomposition.}. 

Let $\mathbf{\tilde{F}}^*=[\mathbf{\tilde{F}}_{\rm c}^*, \mathbf{\tilde{F}}_{\rm r}^*]$ denote the optimal solution for (\ref{eq7}). From (\ref{eq7.3}), we have 
\begin{equation}
\begin{aligned}
  &\log \det ({\sigma ^2}{{\mathbf{I}}_{{N_{{\text{rx}}}}}} + {\mathbf{\tilde H}}{{{\mathbf{\tilde F}}}_{\text{r}}}{\mathbf{\tilde F}}_{\text{r}}^H{{{\mathbf{\tilde H}}}^H}) \hfill \\ 
  = &\log \det ({\sigma ^2}{{\mathbf{I}}_{{N_{{\text{tx}}}} - d}} + {\mathbf{\tilde F}}_{\text{r}}^H{{{\mathbf{\tilde H}}}^H}{\mathbf{\tilde H}}{{{\mathbf{\tilde F}}}_{\text{r}}}) \hfill \\
   = &\log \det ({\mathbf{\tilde F}}_{\text{r}}^H{{\mathbf{H}}_e}{{{\mathbf{\tilde F}}}_{\text{r}}}) \geqslant \log (\prod\limits_{m = 1}^{{N_{{\text{tx}}}} - d} {{\lambda _m}({{\mathbf{H}}_e})} ) \hfill \\ 
\end{aligned} 
\end{equation}
where the first equality is again due to $\det ({\mathbf{I}} + {\mathbf{AB}}) = \det ({\mathbf{I}} + {\mathbf{BA}})$. We can then conclude that the optimal value for problem (\ref{eq7.2}) is given by $\log(\prod\limits_{m = 1}^{{N_{\rm tx}} - d} {{\lambda _m}({{\mathbf{H}}_e})} )$, and the optimal solution ${\mathbf{\tilde F}}_{\rm r}^ * $ is composed of the last $({N_{\rm tx} - d})$ columns of the eigenvector matrix of $\mathbf{H}_e$, i.e., ${\mathbf{V}}$  by (\ref{eq8}). According to the constraint (\ref{eq7}b), the optimal ${\mathbf{\tilde F}}_{\rm c}^ * $ should be orthogonal to ${\mathbf{\tilde F}}_{\rm r}^ * $; i.e., ${\mathbf{\tilde F}}_{\rm c}^ * $ should be in the null space of ${\mathbf{\tilde F}}_{\rm r}^ * $. It then readily follows that ${\mathbf{\tilde F}}_{\rm c}^ * $ consists of the first $d$ columns of ${\mathbf{V}}$; hence, ${{\mathbf{\tilde F}}^*}={\mathbf{V}}$. This in turn leads to  ${{\mathbf{F}}^*} = {\mathbf{L}}{\mathbf{V}}$.   $\hfill\blacksquare$ 

\vspace{2 mm}
\emph{\textbf{Remark 1:}} Consider the MIMO rate maximum problem in (\ref{eq7}). It is well known that the MIMO channel ${\mathbf{\tilde H}} \in {\mathbb{C}^{{N_{{\text{rx}}}} \times {N_{{\text{tx}}}}}}$ can be decomposed into a set of $N_{\rm rx}$ parallel eigen-channels~\cite{14} and there exists an $(N_{\rm tx}-N_{\rm rx})$-dimensional null space of ${\mathbf{\tilde H}}$. If the transmit covariance constraint (\ref{eq7}b) is absent, clearly, the $d$ best eigen-channels should be selected for communications to achieve the maximum sum-rate. Interestingly, \textbf{Theorem 1} leverages Cauchy’s interlace theorem to rigorously prove that this strategy also maximizes the achievable sum-rate, even in the presence of 
transmit covariance constraint (\ref{eq7}b). To this end, 
 the first $d$ columns of the right singular matrix ${\mathbf{V}}$, which corresponds to the $d$ best eigen-channels of ${\mathbf{\tilde H}}$, are selected as the communication beamformers in $\mathbf{\tilde{F}}_{\rm c}^*$.  

To enable the desired transmit covariance ${{\mathbf{R}}_{\rm{des}}}$, another $(N_t-d)$ radar beamformers are placed in the $(N_r-d)$ eigen-channels of ${\mathbf{\tilde H}}$ and $(N_t-N_r)$-dimensional null space of ${\mathbf{\tilde H}}$. Clearly, the radar beamformers packed into the null space of ${\mathbf{\tilde H}}$ would not interfere with the communications at all, whereas the ones in the $(N_r-d)$ eigen-channels of ${\mathbf{\tilde H}}$ can also be made orthogonal to the communication signals with appropriate transceiver processing. As the number $N_r$ increases, more radar beamformers  need to be nullified at the receiver, leading to a higher system complexity; yet, the achievable sum-rate can also be improved since more receive antennas provide greater diversity gain, as will be corroborated by simulations in Section V.$\hfill\blacksquare$

With the optimal $\tilde{\mathbf{F}}^*={\mathbf{V}}$ for (\ref{eq7}), the optimal transmit beamforming matrix for the original problem (\ref{eq5}) is given by $\mathbf{F}^* = \mathbf{LV}$, as stated in \textbf{Theorem 1}. To achieve the corresponding maximum user rate, the optimal receive beamforming matrix at the user should take the classic MMSE form\cite{14}, as given by
\begin{equation} \label{eq8a}
{\mathbf{G}}^{\rm mmse}
= {({{\mathbf{H}}}{{\mathbf{R}}_{\rm{des}}}{{\mathbf{H}}^H} + {\sigma ^2}{{\mathbf{I}}_{{N_{\rm rx}}}})^{ - 1}}{{\mathbf{H}}}{{\mathbf{F}}^*}.
\end{equation}

As shown in (\ref{eq7.1}) and (\ref{eq8a}), both the optimal transmit and receive beamforming solutions are derived in closed form for the single-user MIMO DFRC system. With CSI, i.e., $\mathbf{H}$, available at the transmitter and receiver, these beamforming matrices can be computed and implemented with very low complexity to maximize the user communication rate while achieving the desired beampattern to ensure a stringent sensing performance.

\section{Beamforming Design in Multi-User Case}
We next consider beamforming design in the general multi-user case. Based on (\ref{eq3}) and (\ref{eq3a}), the intended optimization problem can be expressed as
\begin{subequations} \label{eq9}
\begin{align}
   \mathop {\max }\limits_{\mathbf{F}}  & \sum\limits_{k = 1}^K {{\omega_k}} \log \det ({{\mathbf{I}}_d} + {\mathbf{F}}_k^H{\mathbf{H}}_k^H 
  ({\sigma ^2} {{\mathbf{I}}_{{N_{\rm rx}}}}    
   +  \notag \\ 
   &\sum\limits_{i \ne k}^{} {{{\mathbf{H}}_k}{{\mathbf{F}}_i}{\mathbf{F}}_i^H{\mathbf{H}}_k^H}+ {{\mathbf{H}}_k}{{\mathbf{F}}_{\rm r}}{\mathbf{F}}_{\rm r}^H{\mathbf{H}}_k^H{)^{ - 1}}{{\mathbf{H}}_k}{{\mathbf{F}}_k}) \\
  {\text{s}}{\text{.t}}{\text{.   }} &{\mathbf{F}}{{\mathbf{F}}^H} = {{\mathbf{R}}_{\rm{des}}}.
\end{align}
\end{subequations}
Again, the total transmit power constraint for $\mathbf{F}$ is incorporated into the transmit covariance constraint (\ref{eq9}b).  

The problem (\ref{eq9}) is clearly non-convex. Yet, notice that constraint (\ref{eq9}b) indeed confines beamforming design in a complex Grassmann manifold. Hence, one can develop a Riemannian gradient descent-based method to approximately solve (\ref{eq9}) using the existing  \emph{Manopt} toolbox\cite{toolbox}; see details in Appendix~A. 

Different from the above standard manifold optimization approach, we next rely on a judicious reformulation to propose a more efficient BCD algorithm to solve (\ref{eq9}) with better performance.

\subsection{Problem Reformulation}
To make the difficult problem (\ref{eq9}) more tractable, we exploit the connection between weighted sum-rate and weighted MMSE as revealed in \cite{cioffi,wmmse} to reformulate it as follows. 

Recall that $\mathbf{G}= [\mathbf{G}_1, \ldots, \mathbf{G}_K]$ collects the receive beamforming matrices. We define the MSE matrix of user $k$ as 
\begin{equation}  \label{eq9.1}
\begin{aligned}
  &{{\mathbf{E}}_k}({\mathbf{G}},{\mathbf{F}}) = \mathbb{E}\{ ({{{\mathbf{\hat c}}}_k}(t) - {{\mathbf{c}}_k}(t)){({{{\mathbf{\hat c}}}_k}(t) - {{\mathbf{c}}_k}(t))^H}\}   \\
  &\mathop  =  {{\mathbf{I}}_d} - 2\operatorname{Re} \{ {\mathbf{G}}_k^H{{\mathbf{H}}_k}{{\mathbf{F}}_k}\}  + \sum\limits_{i = 1}^K {{\mathbf{G}}_k^H{{\mathbf{H}}_k}{{\mathbf{F}}_i}{\mathbf{F}}_i^H{\mathbf{H}}_k^H{{\mathbf{G}}_k}}   \\
  & \quad + {\mathbf{G}}_k^H{{\mathbf{H}}_k}{{\mathbf{F}}_{\rm r}}{\mathbf{F}}_{\rm r}^H{\mathbf{H}}_k^H{{\mathbf{G}}_k} + {\sigma ^2}{\mathbf{G}}_k^H{{\mathbf{G}}_k}  \\
  &\mathop  =  {{\mathbf{I}}_d} - 2\operatorname{Re} \{ {\mathbf{G}}_k^H{{\mathbf{H}}_k}{{\mathbf{F}}_k}\}  + {\mathbf{G}}_k^H{{\mathbf{H}}_k}{{\mathbf{R}}_{\rm{des}}}{\mathbf{H}}_k^H{{\mathbf{G}}_k} + {\sigma ^2}{\mathbf{G}}_k^H{{\mathbf{G}}_k}, 
\end{aligned} 
\end{equation}
where the last equality holds due to constraint (\ref{eq9}b). Then, we can transform the original problem into a matrix-weighted sum-MSE minimization problem:
\vspace{-0.5em}
\begin{subequations}  \label{eq10}
\begin{align}
  \mathop {\min }\limits_{{\mathbf{F}},{\mathbf{G}},{\mathbf{W}}} &{\text{ }}\sum\limits_{k = 1}^K {{\omega_k}{\text{(Tr}}({{\mathbf{W}}_k}{{\mathbf{E}}_k}) - \log \det ({{\mathbf{W}}_k})} )  \\
  {\text{s}}{\text{.t}}{\text{.   }}&{\mathbf{F}}{{\mathbf{F}}^H} = {{\mathbf{R}}_{\rm{des}}},
\end{align}
\end{subequations}
where the weight matrices ${{\mathbf{W}}_k} \succeq {\mathbf{0}}$, $\forall k$, are introduced as auxiliary optimization variables. By mimicking the relevant proof in \cite{wmmse}, we can establish the following lemma. 

\emph{\textbf{Lemma 2:} 
The problems (\ref{eq9}) and (\ref{eq10}) are equivalent in the sense that the optimal transmit beamforming solution ${{\mathbf{F}}^*}$ for problems (\ref{eq9}) and (\ref{eq10}) are identical; more generally, ${{\mathbf{F}}^*}$ is a stationary point solution for (\ref{eq9})  if and only if it is part of a stationary point solution for (\ref{eq10}). }

\emph{Proof:} See Appendix B. $\hfill\blacksquare$ 

By the equivalence between (\ref{eq9}) and (\ref{eq10}) stated in \textbf{Lemma~2}, we next consider solving the more tractable (\ref{eq10}) using the BCD method.

\subsection{BCD Method}
Note that the optimization variables in problem (\ref{eq10}) consist of ${\mathbf{F}}$, ${\mathbf{G}}$, and ${\mathbf{W}}$. A BCD  based approach can be used to optimize ${\mathbf{G}}$, ${\mathbf{W}}$, and ${\mathbf{F}}$ alternately as follows. 

1) Optimization of $\mathbf{G}$:
Given the fixed ${\mathbf{F}}$, the optimal receive beamformer is clearly provided by the following MMSE one [cf. (\ref{eq8a})]:
\begin{equation} \label{eq11}
\mathbf{G}_k^{\text{mmse}}
= {({{\mathbf{H}}_k}{{\mathbf{R}}_{\rm{des}}}{{\mathbf{H}}_k^H} + {\sigma ^2}{{\mathbf{I}}_{{N_{\rm rx}}}})^{ - 1}}{{\mathbf{H}}_k}{{\mathbf{F}}_k}, {\text{ }}\forall k.
\end{equation}

2) Optimization of $\mathbf{W}$:
By substituting \eqref{eq11} into \eqref{eq9.1}, the corresponding MSE matrix regarding user $k$ is given by
\begin{equation}  \label{eq11a}
\begin{aligned}
 {\mathbf{E}}_k^{{\rm mmse}} = {{\mathbf{I}}_d} - {\mathbf{F}}_k^H{\mathbf{H}}_k^H{({{\mathbf{H}}_k}{{\mathbf{R}}_{\rm{des}}}{\mathbf{H}}_k^H + {\sigma ^2}{{\mathbf{I}}_{{N_{{\text{rx}}}}}})^{ - 1}}{{\mathbf{H}}_k}{{\mathbf{F}}_k}.
\end{aligned} 
\end{equation}
With ${\mathbf{F}}$ and ${\mathbf{G}}={\mathbf{G}^{\text{mmse}}}$ fixed, we have an unconstrained convex problem with ${\mathbf{W}}$. Using the first-order optimality condition, the optimal solution is given by
\begin{equation} \label{eq12}
{\mathbf{W}}_k^* = {({\mathbf{E}}_k^{{\text{mmse}}})^{ - 1}},{\text{ }} \forall k.
\end{equation}


3) Optimization of $\mathbf{F}$:
The most important step of our BCD approach is optimizing the transmit beamforming matrix ${\mathbf{F}}$. With ${\mathbf{G}}$ and $ {\mathbf{W}}$ fixed, problem (\ref{eq10}) becomes
\begin{subequations}  \label{eq13}
\begin{align}
  \mathop {\max }\limits_{\mathbf{F}} {\text{ }}& \sum\limits_{k = 1}^K {{\omega_k}{\text{Re\{Tr}}({{\mathbf{W}}_k}{\mathbf{G}}_k^H{{\mathbf{H}}_k}{{\mathbf{F}}_k})} \}   \\
  {\text{s}}{\text{.t}}{\text{.   }}& {\mathbf{F}}{{\mathbf{F}}^H} = {{\mathbf{R}}_{\rm{des}}}.
\end{align}
\end{subequations}
Since (\ref{eq13}b) is a non-convex quadratic equality constraint, the problem (\ref{eq13}) is non-convex. This is different from the problem studied in \cite{wmmse}, where the transmit beamforming design subproblem is convex and can be solved using general convex optimization methods. However, we next show that such a seemingly difficult non-convex program can be reformulated as an orthogonal Procrustes problem (OPP), which admits a globally optimal solution in closed form.

\subsection{Optimal Transmit Beamforming in Closed Form}
As in the single-user case, we use Cholesky decomposition ${{\mathbf{R}}_{\rm{des}}} = {\mathbf{L}}{{\mathbf{L}}^H}$, and define ${\mathbf{\tilde F}} = {{\mathbf{L}}^{ - 1}}{\mathbf{F}}$ and ${\mathbf{\tilde H}_k} = {\mathbf{H}_k \mathbf{L}}$. Perform the following rearrangements
\begin{equation} 
 \begin{gathered}
  \quad \sum\limits_{k = 1}^K {{\omega _k}{\text{Re}}\{ {\text{Tr}}({{\mathbf{W}}_k}{\mathbf{G}}_k^H{{\mathbf{H}}_k}{{\mathbf{F}}_k})} \}    \hfill \\
     = \sum\limits_{k = 1}^K {{\omega _k}{\text{Re}}\{ {\text{Tr}}({{\mathbf{W}}_k}{\mathbf{G}}_k^H{{{\mathbf{\tilde H}}}_k}{{{\mathbf{\tilde F}}}_k})}\}   \hfill \\
   ={\text{Re}}\{ {\text{Tr}}({\left[ {{\omega _1}{\mathbf{\tilde H}}_1^H{{\mathbf{G}}_1}{\mathbf{W}}_1^H, \cdots ,{\omega _k}{\mathbf{\tilde H}}_K^H{{\mathbf{G}}_K}{\mathbf{W}}_K^H} \right]^H}{{{\mathbf{\tilde F}}}_{\rm c}} )\}, \hfill \\ 
\end{gathered} 
\end{equation} 
and define ${{\mathbf{M}}} = {\left[ {{\omega _1}{\mathbf{\tilde H}}_1^H{{\mathbf{G}}_1}{\mathbf{W}}_1^H, \cdots ,{\omega _k}{\mathbf{\tilde H}}_K^H{{\mathbf{G}}_K}{\mathbf{W}}_K^H} \right]}$. We can then rewrite problem (\ref{eq13}) as
\begin{subequations}  \label{eq14}
\begin{align}
  \mathop {\max }\limits_{{{{\mathbf{\tilde F}}}_{\rm c}}} &{\text{ Re\{Tr}}({\mathbf{M}^H}{{{\mathbf{\tilde F}}}_{\rm c}})\}  \\
  {\text{s}}{\text{.t}}{\text{. }} &{{{\mathbf{\tilde F}}}_{\rm c}}^H{{{\mathbf{\tilde F}}}_{\rm c}} = {{\mathbf{I}}_D} .
\end{align}
\end{subequations}
Interestingly, the non-convex problem (\ref{eq14}) admits a closed-form optimal solution, as inferred by the following lemma.

\emph{\textbf{Lemma 3}\textbf{:} The problem (\ref{eq14}) can be equivalently reformulated into the following OPP}
\begin{subequations} \label{eq15}
\begin{align}
  \mathop {\min }\limits_{{{{\mathbf{\tilde F}}}_{\rm c}}} & {\text{ }}||{{\mathbf{M}}} - {{{\mathbf{\tilde F}}}_{\rm c}}||_F^2  \\
  {\text{s}}{\text{.t}}{\text{. }} &{{{\mathbf{\tilde F}}}_{\rm c}}^H{{{\mathbf{\tilde F}}}_{\rm c}} = {{\mathbf{I}}_D} .
\end{align}
\end{subequations}

\emph{Proof:} Note that 
\begin{equation}
   ||{\mathbf{M}} - {{\mathbf{\tilde F}}_{\text{c}}}||_F^2 ={\text{Tr}}({{\mathbf{M}}^H}{\mathbf{M}} + {{\mathbf{\tilde F}}_{\text{c}}}^H{{\mathbf{\tilde F}}_{\text{c}}}- 2\operatorname{Re} \{ {{\mathbf{M}}^H}{{\mathbf{\tilde F}}_{\text{c}}}\} ).  
\end{equation}
With $\mathbf{G}$ and $\mathbf{W}$ fixed, $\text{Tr}({{\mathbf{M}}^H}{\mathbf{M}})$ is a constant independent of ${{\mathbf{\tilde F}}_{\text{c}}}$.  By the constraint (30b), we also have $\text{Tr}({{\mathbf{\tilde F}}_{\text{c}}}^H{{\mathbf{\tilde F}}_{\text{c}}}) = D$. It then follows that minimizing $||{{\mathbf{M}}} - {{{\mathbf{\tilde F}}}_{\rm c}}||_F^2$ in (\ref{eq15}a) is equivalent to maximizing ${\text{ Re\{Tr}}({\mathbf{M}^H}{{{\mathbf{\tilde F}}}_{\rm c}})\}$ in (\ref{eq14}a). Hence, problem (\ref{eq14}) can be equivalently reformulated into (\ref{eq15}).   $\hfill\blacksquare$

By \cite[Proposition 7]{16}, the OPP (\ref{eq15}) has a unique globally optimal solution in closed form. Specifically, perform the SVD ${{\mathbf{M}}} = {{\mathbf{U}}_{\rm M}}{{\mathbf{\Sigma }}_{\rm M}}{\mathbf{V}}_{\rm M}^H$, with the eigenvalues arranged in descending order along the diagonal of ${{\mathbf{\Sigma }}_{\rm M}}$. Then the unique globally optimal solution for problem (\ref{eq15}) is given by
\begin{equation} \label{eq16}
{\mathbf{\tilde F}}_{\rm c}^* = {{\mathbf{U}}_{\text{M}}}(1:D){\mathbf{V}}_{\text{M}}^H,
\end{equation}
where ${{\mathbf{U}}_{\text{M}}}(1:D)$ collects the first $D$  columns of ${{\mathbf{U}}_{\rm M}}$. 

By the equivalence between problem (\ref{eq14}) and problem (\ref{eq15}), the optimal ${{\mathbf{\tilde F}}_{\text{c}}}$ in (\ref{eq16}) is also the optimal solution for (\ref{eq14}). Based on  (\ref{eq13}b), the optimal radar beamformer ${\mathbf{\tilde F}}_{\rm r}^*$ must be in the null space of ${\mathbf{\tilde F}}_{\rm c}^*$. It then follows
\begin{equation}  \label{eq17}
{\mathbf{\tilde F}}_{\rm r}^* = {{\mathbf{U}}_{\text{M}}}(D + 1:{N_{\rm tx}}),
\end{equation}
where ${{\mathbf{U}}_{\text{M}}}(D + 1:{N_{\rm tx}})$ collects the last $(N_{\rm tx}-D)$ column vectors of ${{\mathbf{U}}_{\rm M}}$. With ${{\mathbf{\tilde F}}^*}=[{\mathbf{\tilde F}}_{\rm c}^*,{\mathbf{\tilde F}}_{\rm r}^*]$, we can in turn obtain the optimal ${{\mathbf{F}}^*} = {\mathbf{L}}{{\mathbf{\tilde F}}^*}$.

\emph{\textbf{Remark 2:}} 
Note that problem (\ref{eq13}) can also be equivalently reformulated into a convex semi-definite program (SDP) by allowing ${\mathbf{F}_{\rm r}}$ to consist of $N_{\rm tx}$ (instead of $N_{\rm tx}-D$) radar beamformers~\cite{12}. General-purpose convex solvers, such as the interior-point method, can be then employed to iteratively compute the optimal $\mathbf{F}^*$ with the complexity $\mathcal{O}({(DN_{\rm tx})^{3.5}}\log (1/\varepsilon ))$, where $\varepsilon $ is the desired accuracy. Such a convex optimization approach is enabled at the cost of additional $D = d \times K$ radar beamforming vectors, which clearly leads to increased implementation complexity, especially when the number $K$ of users is large. Nevertheless, we can show that the additional computational and implementation complexity of the SDP approach [12] cannot bring additional benefit in improving the weighted sum-rate over the proposed closed-form solution for problem (\ref{eq13}); see Appendix C. This is also verified numerically in Section V. $\hfill\blacksquare$

\subsection{Analysis of the Overall Algorithm}
The overall procedure of solving (\ref{eq10})  is summarized in Algorithm~1, with a guaranteed performance stated in the following proposition:

\emph{Proposition 1:  Algorithm 1  surely converges to at least a stationary solution ${{\mathbf{F}}^*}$ of problem (\ref{eq9}). }

\emph{Proof:} The convergence of the algorithm is guaranteed since each BCD iteration monotonically decreases (the lower-bounded) objective function of problem (\ref{eq10}) over a compact feasible set. Moreover, problem (\ref{eq10}) features a differentiable objective function and a separable feasible set, i.e., the overall feasible set ${\cal F(\mathbf{F}, \mathbf{G}, \mathbf{W})={\cal F}(\mathbf{F}) \times {\cal F}(\mathbf{G}) \times {\cal F}(\mathbf{W})}$. It then follows from \cite{BCD} that the BCD-based Algorithm1  surely converges to at least a stationary point solution $({{\mathbf{F}}^*},{{\mathbf{G}}^*},{{\mathbf{W}}^*})$ of problem (\ref{eq10}). 
 
By \textbf{Lemma 2}, the transmit beamforming matrix  ${{\mathbf{F}}^*}$ in the stationary point solution $({{\mathbf{F}}^*},{{\mathbf{G}}^*},{{\mathbf{W}}^*})$ obtained for problem (\ref{eq10}) is also a stationary point solution of problem~(\ref{eq9}). 
$\hfill\blacksquare$

With transmit beamforming matrix $\mathbf{F}^*$ yielded by Algorithm~1, again the optimal receive beamforming matrix at the user $k$ should take the classic MMSE form, as given by [cf. (\ref{eq11})]
\begin{equation}
\mathbf{G}_k^{\text{mmse}}
= {({{\mathbf{H}}_k}{{\mathbf{R}}_{\rm{des}}}{{\mathbf{H}}_k^H} + {\sigma ^2}{{\mathbf{I}}_{{N_{\rm rx}}}})^{ - 1}}{{\mathbf{H}}_k}{{\mathbf{F}}_k^*}, {\text{ }}\forall k.
\end{equation}

Note that at each BCD step, we attain a unique closed-form solution for $\mathbf{F}$, $\mathbf{G}$, and $\mathbf{W}$. Numerical results in Section V also show that the BCD-based Algorithm 1 can converge in only a few iterations and yield a beamforming solution with much better performance than that by the manifold optimization (i.e., Riemannian gradient descent) approach.  With CSI available, the proposed transmit and receive beamforming schemes can be then computed and implemented with low
complexity and high performance.

\begin{algorithm}[H]
\caption{The proposed BCD algorithm for solving (\ref{eq10}).}\label{alg:alg1}
\begin{algorithmic}[1]
\STATE Initialize ${{\mathbf{F}}}={{\mathbf{L}}}$.
\STATE \textbf{Repeat } 
\STATE \hspace{0.5cm} Obtain ${\mathbf{G}}_k$ by (\ref{eq11}), $\forall k$.
\STATE \hspace{0.5cm} Obtain ${\mathbf{W}}_k$ by (\ref{eq12}), $\forall k$.
\STATE \hspace{0.5cm} Obtain ${\mathbf{\tilde F}}$  using (\ref{eq16}) and (\ref{eq17}) , and update ${{\mathbf{F}}} = {\mathbf{L}}{{\mathbf{\tilde F}}}$.
\STATE \textbf{Until} convergence.
\end{algorithmic}
\label{alg1}
\end{algorithm}

\section{Numerical Results}

In this section, we carry out extensive Monte-Carlo simulations to gauge the performance of the proposed schemes. Suppose that the BS and the users are equipped with uniform linear arrays with half-wavelength spacing between adjacent antennas. The default number of transmit antennas at the BS is $N_{\rm tx}=16$. Unless otherwise specified, assume that the same number of receive antennas are used at the BS for radar echo signals. Assume a Rayleigh fading model where each element of channel matrix $\mathbf{H}_k,\, \forall k$,  is independently generated according to the standard complex Gaussian distribution $\mathcal{CN}(0,1)$. While the Rayleigh fading model is used in simulations, our proposed schemes apply to any other fading models. The total transmit power budget at the BS is $P$, and the transmit signal-to-noise-ratio (SNR) is $P/{\sigma ^2}$  (as full power transmissions are always performed). 

Without loss of generality, we assume that the radar system directs its beams towards $J = 3$ targets of interest located at ${\theta _1} =  - {60^ \circ }$, ${\theta _2} = {0^ \circ }$, and ${\theta _3} = {60^ \circ }$ with the complex reflection coefficient set to 1. It is also assumed that only the targets reflect the transmitted signals. The echo signal is corrupted by zero-mean Gaussian noise with variance $\sigma_r^2$, and the radar transmit SNR is $P/{\sigma_r^2}$. 

The ideal beampattern consisting of three main beams with a beam width of  $\Delta  = 9^\circ $ each is given by
\begin{equation} \label{eq19a}
{\tilde \pi_d}(\theta ) = \left\{ {\begin{array}{lc}
  {1,\quad \text{if }{\theta _j} \!-\! \frac{\Delta}{2} \!\leqslant \!\theta  \leqslant {\theta _j}\!\! +\!\! \frac{\Delta}{2},\,j \!\!= \!\!1,2,3;} \\ 
  {0,\quad {\text{otherwise}}.} 
\end{array}} \right.
\end{equation}
The desired transmit covariance  ${{\mathbf{R}}_{\rm{des}}}$ is determined by solving a constrained least-squares problem to minimize the radar beampattern matching errors, as in \cite{2008tae}.

To compare with the ``Derived optimal" solution in Theorem~1 for the single-user case and ``Proposed BCD" solution yielded by Algorithm 1, we consider the following benchmark schemes:
\begin{itemize}
  \item 
  \emph{DPC} \cite{12}: DPC is employed for the sum-rate maximization problem to pre-cancel the interference caused by radar and other users' signals at the transmitter. However, the DPC scheme proposed in \cite{12} only applies to MISO systems. For a fair comparison, when user $k$ is equipped with multiple antennas to receive a single data stream, the row with the largest 2-norm in ${{\mathbf{H}}_k}$ is selected for implementation of such a DPC scheme.
   \item 
  \emph{Manopt}\cite{manopt}: Recall that constraint (\ref{eq9}b) defines a complex Grassmann manifold for the feasible set of $\mathbf{F}$. Hence, problem (\ref{eq9}) could be solved by the Riemannian gradient descent method delineated in Appendix A. Treating the single-user problem (\ref{eq5}) as a special case of problem (\ref{eq9}) with $K=1$, this manifold optimization method can be used to approximately solve (\ref{eq5}) as well.
  \item 
\emph{MMSE filter}\cite{11}: Here, MMSE is used as the optimization objective instead of the weighted sum-rate (\ref{eq9}a), and the transmit and receive beamformers are optimized in an alternating manner until convergence.
 \item 
  \emph{Cholesky}:  The  transmit beamforming matrix is set to ${\mathbf{F}} = {\mathbf{L}}$, where ${\mathbf{L}}$ is obtained by Cholesky decomposition of ${{\mathbf{R}}_{\rm{des}}}$ as defined in (\ref{eq6}). Note that ${\mathbf{L}}$ also serves as the initial point of our proposed BCD algorithm.
  
\end{itemize}
Besides, a scheme called ``BCD w/ SDP” referring to Algorithm 1 using the SDP approach   \cite{12} to solve the transmit beamforming subproblem  (\ref{eq13}), is also considered on occasions. Our simulations are based on MATLAB and executed on an Intel Core i7-10700K CPU @ 3.80GHz. 

\subsection{Single-User Case}
\begin{figure}[!t]
\centering
\includegraphics[width=3.5in]{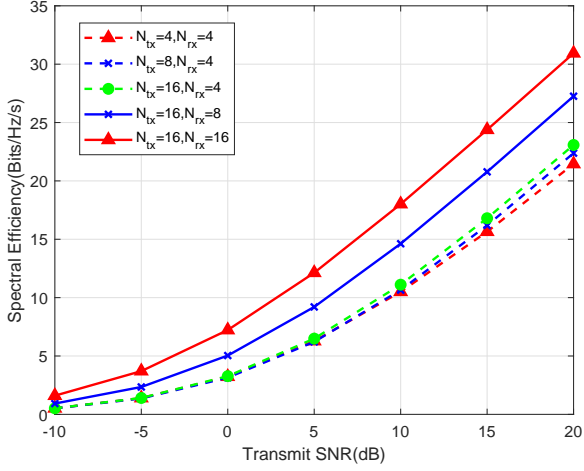}
\caption{Spectral efficiency versus transmit SNR with different numbers of transmit and receive antennas in the single-user case.}
\label{fig_1}
\end{figure}

Suppose that the BS transmits $d=4$ independent data streams to the user. Fig.~\ref{fig_1} depicts the spectral efficiency (i.e., the achievable user rate in (\ref{eq3a}) with the unit bandwidth)  under different transmit SNRs with different numbers of transmit and receive antennas. It is observed that with more receive antennas, the spectral efficiency is improved since higher-order diversity gain is provided. However, spectral efficiency grows marginally as the number of transmit antennas increases. This is because the transmit covariance constraint indeed confines the design space to a  complex Grassmann manifold so that the additional degrees of freedom resulting from an increased number of transmit antennas can hardly be exploited. 

\begin{figure}[!t]
\centering
\includegraphics[width=3.5in]{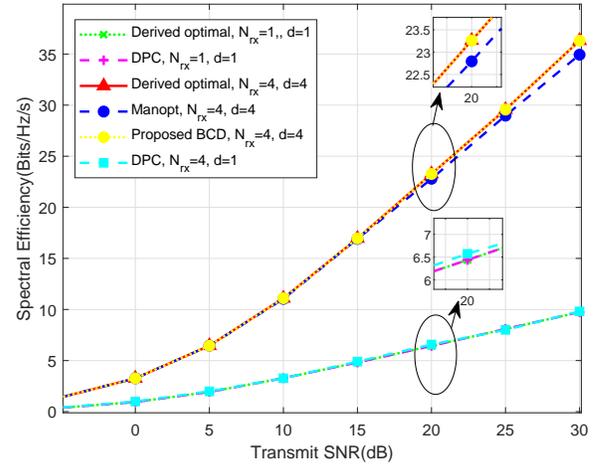}
\caption{Spectral efficiency versus transmit SNR under different schemes in the single-user case.}
\label{fig_1.1}
\end{figure}

Fig.~\ref{fig_1.1} depicts the spectral efficiency versus the transmit SNR under different schemes. With $d=4$ and $N_{\rm rx}=4$, it is observed that the closed-form optimal precoder derived in Theorem 1 outperforms the benchmark Manopt, especially in a high SNR regime, manifesting the suboptimality of the Manopt method. Note that the proposed BCD algorithm can also be used to solve (\ref{eq5}) by treating the single-user problem as a special case of the multi-user one (\ref{eq9}). It is shown that the difference between the achieved spectral efficiency by the BCD solution and the closed-form solution is invisible. This indicates that the proposed low-complexity BCD solution can provide (near-)optimal performance in this single-user case.  With $d=1$ and $N_{\rm rx}=1$, it is also observed that the proposed closed-form solution without DPC shares the same performance as the DPC-based scheme in \cite{12}. 

As revealed in \textbf{Remark 1}, the channel ${\mathbf{\tilde H}}$ can be decomposed into $N_{\rm rx}$ parallel eigen-channels and the interference caused by radar signals can be eliminated with appropriate transmit- and receive-processing. While DPC pre-cancels the interference of radar signals on the transmit side, the proposed optimal solution can achieve the same goal on the receive side without DPC in the single-user case, leading to the same performance. The performance of DPC with $d=1$ and $N_{\rm rx}=4$ is also examined. It shows that little gain is achieved if more receive antennas are equipped since, here, the receiver only performs an antenna-selection-like processing to facilitate the operation of the DPC scheme in \cite{12}. In this case, the MIMO multiplexing and diversity gains cannot be fully exploited. On the other hand, the spectral efficiency of the proposed solution with $d=4$ data streams is 3.5 times greater than that with $d=1$ data stream, confirming that the proposed solution can effectively leverage the MIMO transmissions to benefit the communication capacity of DFRC systems. 

\begin{figure}[!t]
\centering
\includegraphics[width=3.5in]{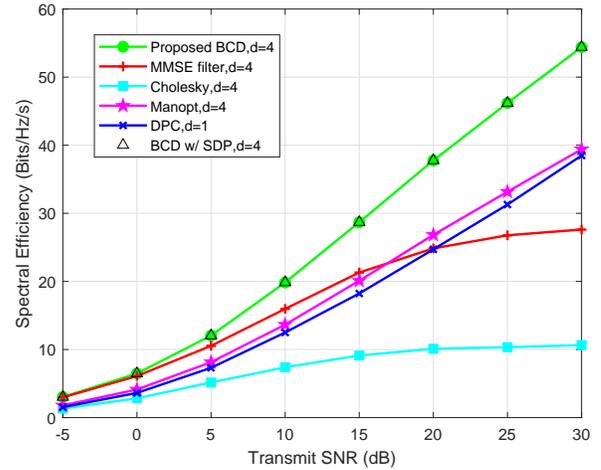}
\caption{Spectral efficiency versus transmit SNR under different schemes in the multi-user case.}
\label{fig_2}
\end{figure}

\subsection{Multi-User Case}
We assume that there exist  $K = 4$ users, each equipped with $N_{\rm rx} = 4$ receive antennas. Without loss of generality,  we set ${\omega_k}=1, \forall k$. In this case, the sum-rate in (\ref{eq3}) again corresponds to spectral efficiency when unit bandwidth is used. Unless otherwise specified, we assume that the BS sends $d=4$ independent data streams to each user in the downlink. 

Fig. \ref{fig_2} shows the spectral efficiency of different beamforming schemes under different transmit SNRs. It is observed that the proposed BCD-based scheme significantly outperforms the four benchmark schemes. The proposed BCD solution performs much better (e.g., yields a 97\% higher special efficiency at 30 dB SNR) than the MMSE filter,  especially in a high SNR regime. This is due to the fact that MMSE does not necessarily lead to the sum-rate maximization in most cases. It is also clearly shown that the proposed BCD scheme exhibits superior performance compared to Manopt, providing a solution with at least 40\% higher spectral efficiency. Although theoretically speaking, both the proposed BCD and the manifold optimization method can achieve stationary point solutions for problem (\ref{eq9}) upon convergence, simulation results verify that the proposed BCD algorithm can take advantage of the derived closed-form solutions for the subproblems in each step to facilitate faster convergence and yield a better-quality overall beamforming design than Manopt. In addition, the big gap (e.g., 5.1 times higher spectral efficiency at 30 dB SNR) between the proposed BCD solution and Cholesky decomposition clearly corroborates the substantial gain from our proposed BCD iterations. Compared with the DPC scheme that only supports $d= 1$ data stream transmission to each user, the proposed BCD solution supporting multi-stream transmissions per user could achieve a much (at least 41\%) higher spectral efficiency. It is worth mentioning that implementation of the DPC scheme could dramatically raise the complexity of the multi-user MIMO systems and might even be computationally prohibitive in practice. It can also be seen that the proposed BCD achieves exactly the same spectral efficiency as the ``BCD w/ SDP'' for which additional $D$ radar beamformers are required, demonstrating numerically that the extra computational and implementation complexity with the SDP approach in \cite{12} is not necessary in our case.

\begin{figure}[!t]
\centering
\includegraphics[width=3.5in]{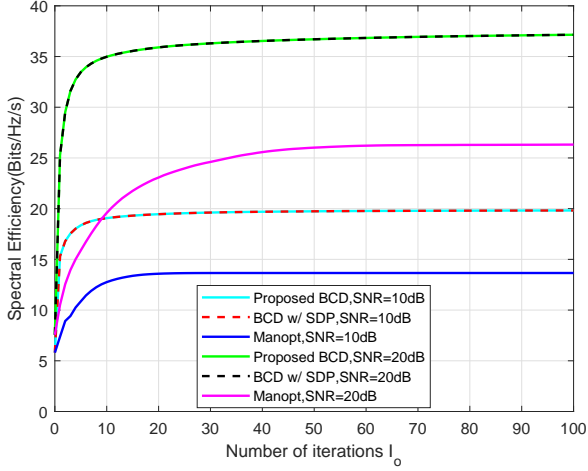}
\caption{Convergence behaviour under different schemes with $N_{\rm tx}=16$, $d=4$, $K=4$, and $N_{\rm rx}=4$.}
\label{fig_3}
\end{figure}

 Fig. \ref{fig_3} shows the convergence behavior of three different methods under different transmit SNRs, where $d=4$, $K=4$, and $N_{\rm rx}=4$. It is evident that the proposed BCD quickly converges within only a few iterations. In particular, the proposed BCD not only converges much faster than the Manopt, but also yields beamforming solutions with much higher spectral efficiency, e.g., 45\% higher in the 10 dB SNR case, and 41\% higher in the 20 dB SNR case, than the Manopt. The ``BCD w/ SDP'' method shares exactly the same convergence process with the proposed BCD. This is not surprising since the SDP approach used there indeed always provides the same communication beamforming solution as with the proposed closed-form solution in each BCD step, as proved in Appendix~C. On the other hand, our simulations also show that the CPU time required for the ``BCD w/ SDP'' method is at least three orders of magnitude higher than that for the proposed BCD method. Overall, compared to the other methods, the computational complexity and implementation time with the proposed BCD approach is very low, justifying its applicability in practical systems.

\begin{figure}[!t]
\centering
\includegraphics[width=3.5in]{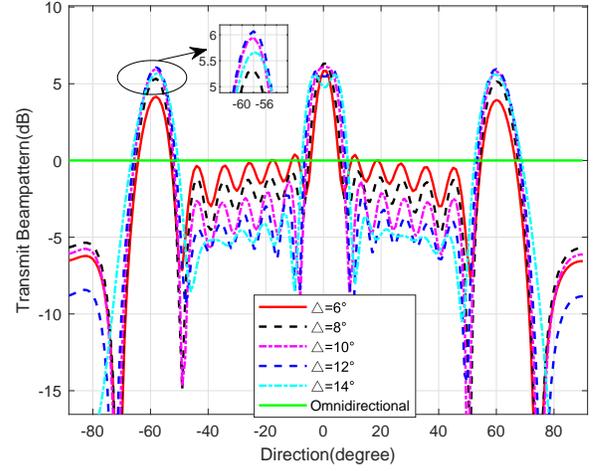}
\caption{Transmit beampattern with different ${{\mathbf{R}}_{\rm{des}}}$.}
\label{fig_5a}
\end{figure}

\begin{figure*}[!t]
\centering
\subfloat[]{\includegraphics[width=2.3in]{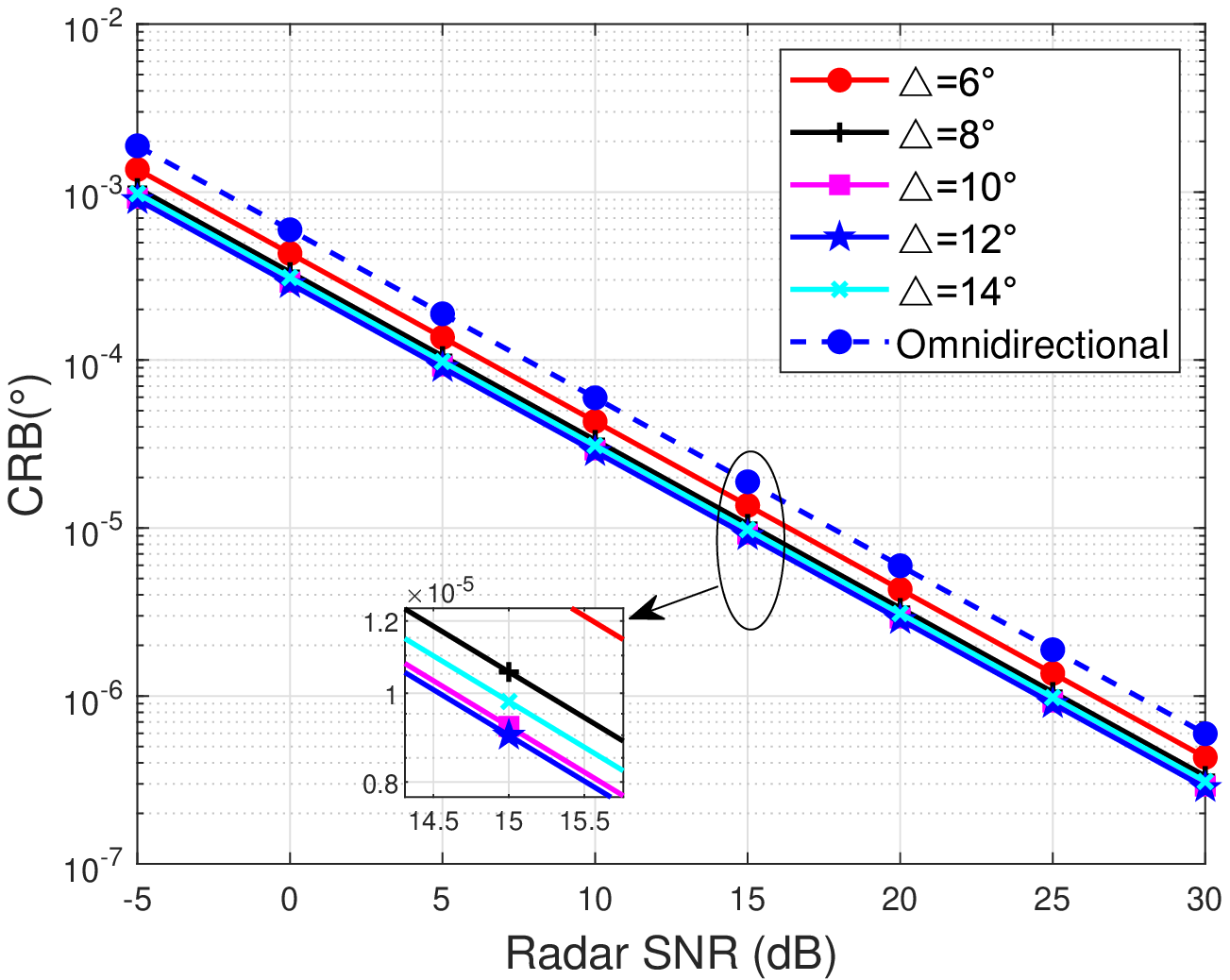}%
\label{fig_first_case}}
\hfil
\subfloat[]{\includegraphics[width=2.3in]{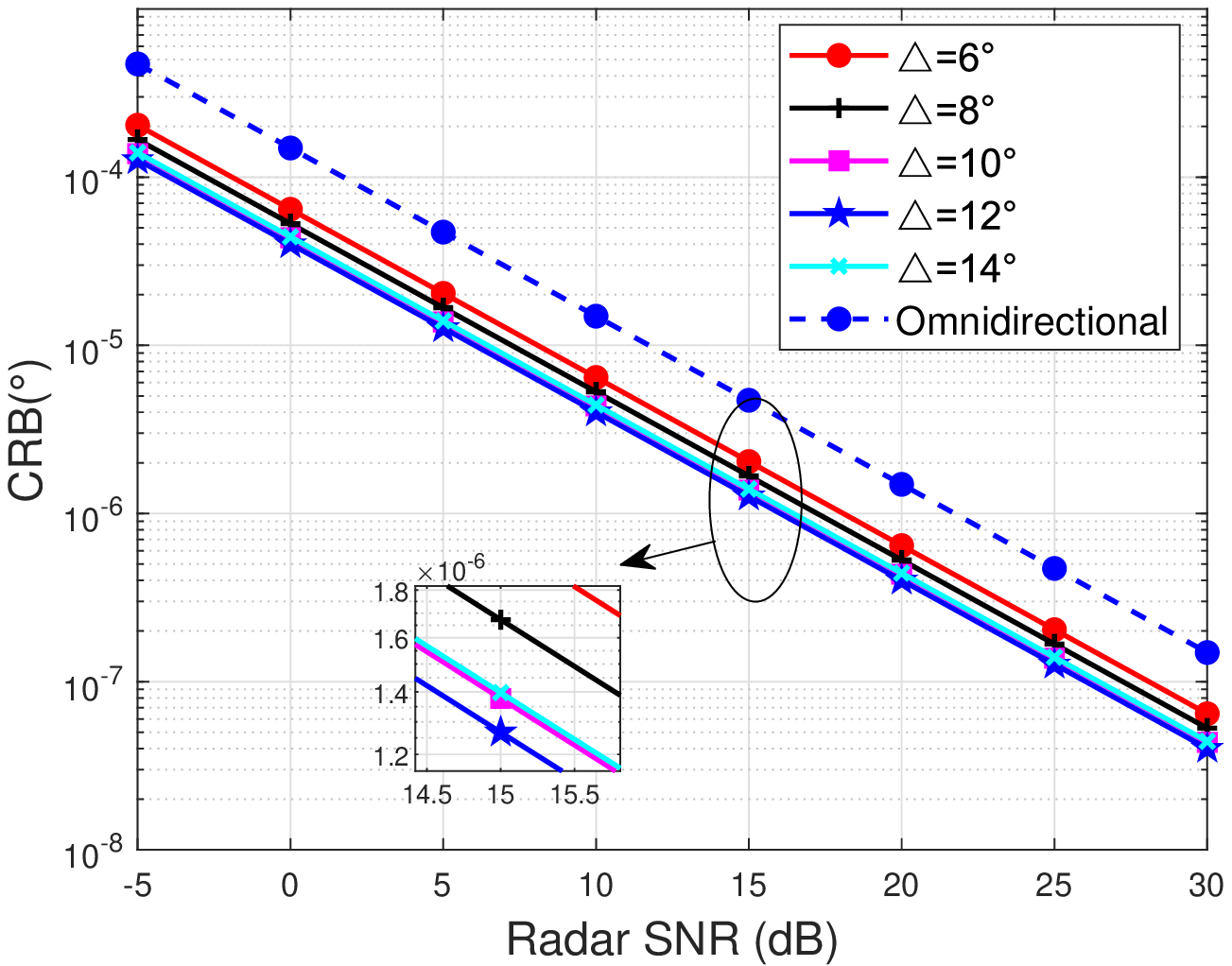}%
\label{fig_second_case}}
\hfil
\subfloat[]{\includegraphics[width=2.3in]{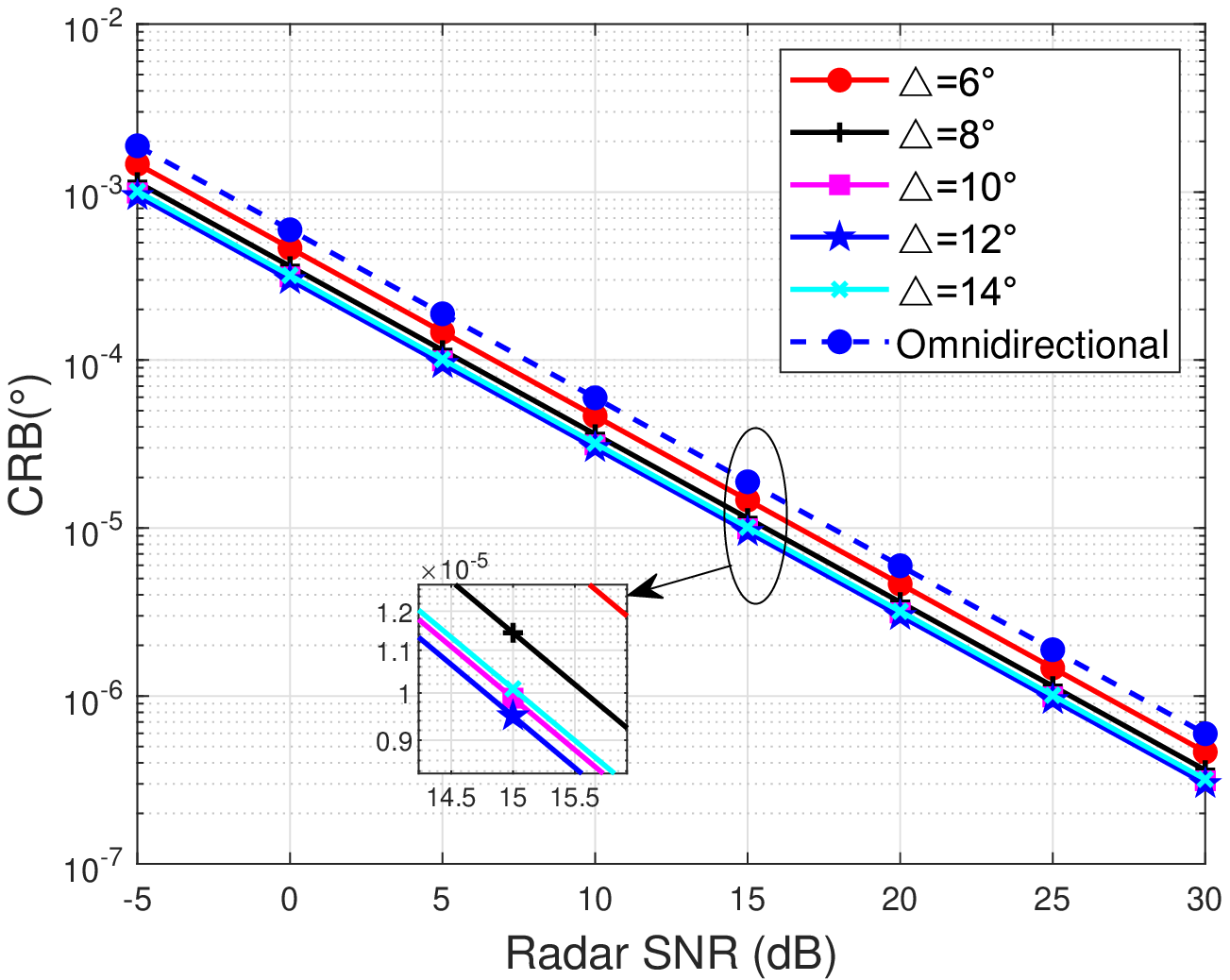}%
\label{fig_third_case}}
\caption{CRB of DOA estimation versus radar SNR under different ${{\mathbf{R}}_{\rm{des}}}$. (a) Target at ${\theta _1} =  - {60^ \circ }$. (b) Target at ${\theta _2} = {0^ \circ }$. (c) Target at ${\theta _3} = {60^ \circ }$.}
\label{fig_6}
\end{figure*}

\begin{figure}[!t]
\centering
\includegraphics[width=3.5in]{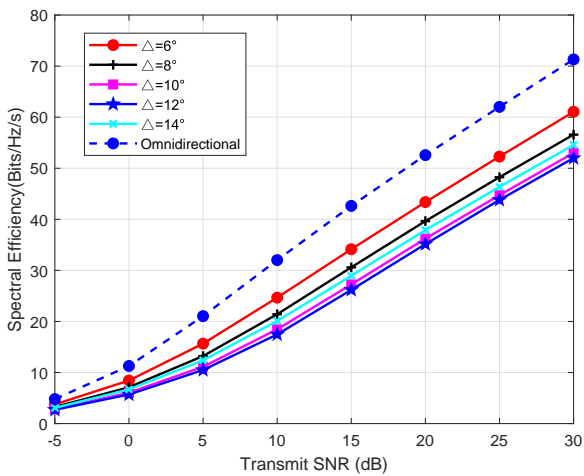}
\caption{Spectral efficiency versus transmit SNR under different ${{\mathbf{R}}_{\rm{des}}}$.}
\label{fig_5}
\end{figure}

\subsection{Trade-off between Sensing and Communication}
In our DFRC system, the desired transmit covariance matrix needs to be achieved for sensing performance guarantee. Consider now that the resultant radar beampattern is used for DOA estimation. The sensing performance can be measured by CRB \cite[Eq.~(6)]{1bit}, which provides a lower bound for the MSE of the intended DOA estimation. To evaluate the performance trade-off between sensing and communication, we adjust the beam width of the mainlobes, i.e., $\Delta$ in (\ref{eq19a}), to yield different ${{\mathbf{R}}_{\rm{des}}}$ matrices. A special case is also investigated, where the radar operating in a searching mode forms the omnidirectional beampattern to coarsely locate the targets; i.e., setting ${{\mathbf{R}}_{\rm{des}}} = \frac{P}{N_{\rm tx}}{{\mathbf{I}}_{N_{\rm tx}}}$ \cite{lijian}.

Figs.~\ref{fig_5a} shows the transmit beampattern with different ${{\mathbf{R}}_{\rm{des}}}$ where we set $P=1$. The sensing performance is highly related to energy directed towards the targets' directions, displayed in the beampattern. Figs.~\ref{fig_6} and~\ref{fig_5} plot the CRB of the DOA estimation under different radar SNRs, and the corresponding spectral efficiency under different transmit SNRs, respectively, where we set $d=4$, $K=4$, and $N_{\rm rx}=4$. It is observed that when $\Delta=12^ \circ$, the DFRC system has the best sensing performance and the worst communication performance. As $\Delta$ increases or decreases, better communication performance is achieved at the cost of certain sensing performance loss. 
As expected, the worst sensing performance and the best communication performance are achieved when the omnidirectional beampattern is maintained, which can be interpreted as the extreme case, i.e., $\Delta=120^\circ$.

\section{Conclusion}
We developed optimal beamforming designs that maximize the weighted sum-rate of communication under a prescribed transmit covariance constraint for MIMO DFRC systems. Leveraging Cauchy's interlace theorem, the globally optimal transmit and receive beamforming solutions were derived in closed form in the single-user case. The proposed approach was then generalized to the multi-user case, where an efficient BCD-based algorithm was proposed to find a high-quality beamforming design with fast convergence and low computational complexity. Extensive simulations corroborated the merits of our beamforming designs in achievable communication rates and sensing performance guarantee, and showed the superiority of the proposed schemes to the existing benchmarks, with at least  40\% higher spectral efficiency under a multi-user MIMO setting in a high SNR regime.

{\appendices
\section{Riemannian Gradient Descent-Based Method for solving (\ref{eq9})}

To exploit the inherent orthogonality among the beamforming vectors as in the single-user case, we use Cholesky decomposition ${{\mathbf{R}}_{\rm{des}}} = {\mathbf{L}}{{\mathbf{L}}^H}$, and define ${\mathbf{\tilde F}} = {{\mathbf{L}}^{ - 1}}{\mathbf{F}}$ and ${\mathbf{\tilde H}_k} = {\mathbf{H}_k \mathbf{L}}$. Problem (\ref{eq9}) can  be equivalently rewritten as 
\begin{subequations}\label{eqa1}
\begin{align}
   \mathop {\max }\limits_{\mathbf{\tilde F}}  & \sum\limits_{k = 1}^K {{\omega_k}} \log \det ({{\mathbf{I}}_d} + {\mathbf{\tilde F}}_k^H{\mathbf{\tilde H}}_k^H 
  ({\sigma ^2} {{\mathbf{I}}_{{N_{\rm rx}}}}    
   +  \notag \\ 
   &\sum\limits_{i \ne k}^{} {{{\mathbf{\tilde H}}_k}{{\mathbf{\tilde F}}_i}{\mathbf{\tilde F}}_i^H{\mathbf{\tilde H}}_k^H}+ {{\mathbf{\tilde H}}_k}{{\mathbf{\tilde F}}_{\rm r}}{\mathbf{\tilde F}}_{\rm r}^H{\mathbf{\tilde H}}_k^H{)^{ - 1}}{{\mathbf{\tilde H}}_k}{{\mathbf{\tilde F}}_k}) \\
  {\text{s}}{\text{.t}}{\text{.   }} &{\mathbf{\tilde F}}{{\mathbf{\tilde F}}^H} = {{\mathbf{I}}_{N_{\rm tx}}}.
\end{align}
\end{subequations}
Constraint (\ref{eqa1}b) actually confines ${\mathbf{\tilde F}}$ on a complex Grassmann manifold (a type of Riemannian manifold). Problem (\ref{eqa1}) could be solved using the Riemannian gradient descent algorithm~\cite{manopt}.  Specifically, the optimization variable ${\mathbf{\tilde F}}$ is iteratively updated in the direction of the Riemannian gradient, which is obtained by projecting the Euclidean conjugate gradient onto the tangent space of a given point on the Riemannian manifold. To implement such an algorithm with the existing \emph{Manopt} toolbox, we only need to derive the conjugate gradient of (\ref{eqa1}a) with respect to ${\mathbf{\tilde F}}$ in the Euclidean space.

To this end, we perform the following manipulation on the achievable rate expression of user $k$  as in (\ref{eqa2}) at the top of the next page, where $(a)$ is due to $\det ({\mathbf{I}} + {\mathbf{AB}}) = \det ({\mathbf{I}} + {\mathbf{BA}})$, $(b)$ is based on constraint (\ref{eqa1}b), and $(c)$ is obtained since $\det ({{\mathbf{B}}^{ - 1}}{\mathbf{A}}) = \det ({\mathbf{A}})/\det ({\mathbf{B}})$.
\begin{figure*} 
\begin{subequations}  \label{eqa2}
\begin{align}
  &\log \det ({{\mathbf{I}}_d} + {\mathbf{\tilde F}}_k^H{\mathbf{\tilde H}}_k^H{({\sigma ^2}{{\mathbf{I}}_{{N_{{\text{rx}}}}}} + \sum\limits_{i \ne k}^{} {{{{\mathbf{\tilde H}}}_k}{{{\mathbf{\tilde F}}}_i}{\mathbf{\tilde F}}_i^H{\mathbf{\tilde H}}_k^H}  + {{{\mathbf{\tilde H}}}_k}{{{\mathbf{\tilde F}}}_{\text{r}}}{\mathbf{\tilde F}}_{\text{r}}^H{\mathbf{\tilde H}}_k^H)^{ - 1}}{{{\mathbf{\tilde H}}}_k}{{{\mathbf{\tilde F}}}_k}) \hfill \\
  \mathop  = \limits^{(a)} &\log \det ({{\mathbf{I}}_{{N_{{\text{rx}}}}}} + {({\sigma ^2}{{\mathbf{I}}_{{N_{{\text{rx}}}}}} + \sum\limits_{i \ne k}^{} {{{{\mathbf{\tilde H}}}_k}{{{\mathbf{\tilde F}}}_i}{\mathbf{\tilde F}}_i^H{\mathbf{\tilde H}}_k^H}  + {{{\mathbf{\tilde H}}}_k}{{{\mathbf{\tilde F}}}_{\text{r}}}{\mathbf{\tilde F}}_{\text{r}}^H{\mathbf{\tilde H}}_k^H)^{ - 1}}{{{\mathbf{\tilde H}}}_k}{{{\mathbf{\tilde F}}}_k}{\mathbf{\tilde F}}_k^H{\mathbf{\tilde H}}_k^H) \hfill \\
   = &\log \det ({({\sigma ^2}{{\mathbf{I}}_{{N_{{\text{rx}}}}}} + \sum\limits_{i \ne k}^{} {{{{\mathbf{\tilde H}}}_k}{{{\mathbf{\tilde F}}}_i}{\mathbf{\tilde F}}_i^H{\mathbf{\tilde H}}_k^H}  + {{{\mathbf{\tilde H}}}_k}{{{\mathbf{\tilde F}}}_{\text{r}}}{\mathbf{\tilde F}}_{\text{r}}^H{\mathbf{\tilde H}}_k^H)^{ - 1}}({\sigma ^2}{{\mathbf{I}}_{{N_{{\text{rx}}}}}} + \sum\limits_{i = 1}^K {{{{\mathbf{\tilde H}}}_k}{{{\mathbf{\tilde F}}}_i}{\mathbf{\tilde F}}_i^H{\mathbf{\tilde H}}_k^H}  + {{{\mathbf{\tilde H}}}_k}{{{\mathbf{\tilde F}}}_{\text{r}}}{\mathbf{\tilde F}}_{\text{r}}^H{\mathbf{\tilde H}}_k^H)) \hfill \\
   = &\log \det ({({\sigma ^2}{{\mathbf{I}}_{{N_{{\text{rx}}}}}} + \sum\limits_{i \ne k}^{} {{{{\mathbf{\tilde H}}}_k}{{{\mathbf{\tilde F}}}_i}{\mathbf{\tilde F}}_i^H{\mathbf{\tilde H}}_k^H}  + {{{\mathbf{\tilde H}}}_k}{{{\mathbf{\tilde F}}}_{\text{r}}}{\mathbf{\tilde F}}_{\text{r}}^H{\mathbf{\tilde H}}_k^H)^{ - 1}}({\sigma ^2}{{\mathbf{I}}_{{N_{{\text{rx}}}}}} + {{{\mathbf{\tilde H}}}_k}{\mathbf{\tilde F}}{{{\mathbf{\tilde F}}}^H}{\mathbf{\tilde H}}_k^H)) \hfill \\
  \mathop  = \limits^{(b)} &\log \det ({({\sigma ^2}{{\mathbf{I}}_{{N_{{\text{rx}}}}}} + \sum\limits_{i \ne k}^{} {{{{\mathbf{\tilde H}}}_k}{{{\mathbf{\tilde F}}}_i}{\mathbf{\tilde F}}_i^H{\mathbf{\tilde H}}_k^H}  + {{{\mathbf{\tilde H}}}_k}{{{\mathbf{\tilde F}}}_{\text{r}}}{\mathbf{\tilde F}}_{\text{r}}^H{\mathbf{\tilde H}}_k^H)^{ - 1}}({\sigma ^2}{{\mathbf{I}}_{{N_{{\text{rx}}}}}} + {{{\mathbf{\tilde H}}}_k}{\mathbf{\tilde H}}_k^H)) \hfill \\
  \mathop  = \limits^{(c)} &\log \det ({\sigma ^2}{{\mathbf{I}}_{{N_{{\text{rx}}}}}} + {{{\mathbf{\tilde H}}}_k}{\mathbf{\tilde H}}_k^H) - \log \det ({\sigma ^2}{{\mathbf{I}}_{{N_{{\text{rx}}}}}} + \sum\limits_{i \ne k}^{} {{{{\mathbf{\tilde H}}}_k}{{{\mathbf{\tilde F}}}_i}{\mathbf{\tilde F}}_i^H{\mathbf{\tilde H}}_k^H}  + {{{\mathbf{\tilde H}}}_k}{{{\mathbf{\tilde F}}}_{\text{r}}}{\mathbf{\tilde F}}_{\text{r}}^H{\mathbf{\tilde H}}_k^H), \hfill 
\end{align} 
\end{subequations}
\hrulefill
\end{figure*} 

As the term $\log \det ({\sigma ^2}{{\mathbf{I}}_{N_{\rm rx}}}{\mathbf{ + \tilde H}}{{{\mathbf{\tilde H}}}^H})$ in (\ref{eqa2}f) is independent of $\mathbf{\tilde F}$, we only need to compute the gradient with respect to $\mathbf{\tilde F}$ for the second term in (\ref{eqa2}f), which is re-defined as 
\begin{equation} \label{eqa3}
   {f_k}({\mathbf{\tilde F}}) =  - \log \det ({\sigma ^2}{{\mathbf{I}}_{{N_{{\text{rx}}}}}} + \sum\limits_{i \ne k}^{} {{{{\mathbf{\tilde H}}}_k}{{{\mathbf{\tilde F}}}_i}{\mathbf{\tilde F}}_i^H{\mathbf{\tilde H}}_k^H}  + {{\mathbf{\tilde H}}_k}{{\mathbf{\tilde F}}_{\text{r}}}{\mathbf{\tilde F}}_{\text{r}}^H{\mathbf{\tilde H}}_k^H)
\end{equation}
For simplicity, we introduce auxiliary variables ${{\mathbf{S}}_k} \in {\mathbb{C}^{{N_{\rm tx}} \times ({N_{\rm tx}} - D)}}$, satisfying
\begin{equation}  \label{eqa4}
{\mathbf{\tilde F}}{{\mathbf{S}}_k} = [{{\mathbf{\tilde F}}_1}, \cdots ,{{\mathbf{\tilde F}}_{k - 1}},{{\mathbf{\tilde F}}_{k + 1}}, \cdots ,{{\mathbf{\tilde F}}_{K,}}{{\mathbf{\tilde F}}_{\rm r}}].
\end{equation}
Clearly ${{\mathbf{S}}_k}$ simply serves as a selection matrix that excludes $\mathbf{\tilde F}_k$ from $\mathbf{\tilde F}$. Substituting (\ref{eqa4}) into (\ref{eqa3}), we immediately have
\begin{equation}   \label{eqa5}
\begin{gathered}
  {f_k}({\mathbf{\tilde F}}) =  - \log \det ({\sigma ^2}{{\mathbf{I}}_{{N_{{\text{rx}}}}}} + {{{\mathbf{\tilde H}}}_k}(\sum\limits_{i \ne k}^{} {{{{\mathbf{\tilde F}}}_i}{\mathbf{\tilde F}}_i^H}  + {{{\mathbf{\tilde F}}}_{\text{r}}}{\mathbf{\tilde F}}_{\text{r}}^H){\mathbf{\tilde H}}_k^H) \hfill \\
   =  - \log \det ({\sigma ^2}{{\mathbf{I}}_{{N_{{\text{rx}}}}}} + {{{\mathbf{\tilde H}}}_k}{\mathbf{\tilde F}}{{\mathbf{S}}_k}{\mathbf{S}}_k^H{{\mathbf{\tilde F}}^H}{\mathbf{\tilde H}}_k^H) .
\end{gathered} 
\end{equation}

Recall that the conjugate gradient of a scalar function ${f_k}({\mathbf{\tilde F}})$ with respect to a complex-valued variable $\mathbf{\tilde F}$, is defined as $\nabla_{\mathbf{\tilde F}} {f_k}({\mathbf{\tilde F}}) = \frac{{\partial {f_k}({\mathbf{\tilde F}})}}{{\partial ({{{\mathbf{\tilde F}}}^ C })}}$\cite{commatrix}. Then, we can show that:

\emph{\textbf{Lemma 4:} The conjugate gradient of ${f_k}({\mathbf{\tilde F}})$  with respect to $\mathbf{\tilde F}$
is given by }
\begin{equation} \label{eqa51}
    \nabla_{\mathbf{\tilde F}} {f_k}({\mathbf{\tilde F}}) = -{\mathbf{\tilde H}}_k^H{({\sigma ^2}{{\mathbf{I}}_{{N_{{\text{rx}}}}}} + {{\mathbf{\tilde H}}_k}{\mathbf{\tilde F}}{{\mathbf{S}}_k}{\mathbf{S}}_k^H{{\mathbf{\tilde F}}^H}{\mathbf{\tilde H}}_k^H)^{ - 1}}{{\mathbf{\tilde H}}_k}{\mathbf{\tilde F}}{{\mathbf{S}}_k}{\mathbf{S}}_k^H.
\end{equation}

\emph{Proof:} Based on basic differentiation rules for
complex-value matrices\cite{commatrix}, the differential of
${f_k}({\mathbf{\tilde F}})$ is given as 
\begin{equation} \label{eqa6}
    \begin{aligned}
  {\rm d}({f_k}({\mathbf{\tilde F}})) =& {\text{Tr}}(\nabla_{\mathbf{\tilde F}} {f_k}{({\mathbf{\tilde F}})^T}{\rm d}({{{\mathbf{\tilde F}}}^ C })) \hfill \\
   =& {\text{Tr}}(\nabla_{\mathbf{\tilde F}} {f_k}({\mathbf{\tilde F}}){\rm d}({{{\mathbf{\tilde F}}}^H})),
    \end{aligned} 
\end{equation}
where ${\rm d}( \cdot)$ denotes the differential operator. The second equality holds due to $\text{Tr}(\mathbf{A}\mathbf{B})=\text{Tr}(\mathbf{B}\mathbf{A})$ and $\text{Tr}(\mathbf{A})=\text{Tr}(\mathbf{A}^T)$. In addition, we have
\begin{equation} \label{eqa7}
\begin{gathered}
  {\rm d}({f_k}({\mathbf{\tilde F}})) =  - {\rm d}(\log \det ({\sigma ^2}{{\mathbf{I}}_{{N_{{\text{rx}}}}}} + {{{\mathbf{\tilde H}}}_k}{\mathbf{\tilde F}}{{\mathbf{S}}_k}{\mathbf{S}}_k^H{{{\mathbf{\tilde F}}}^H}{\mathbf{\tilde H}}_k^H)) \hfill \\
  \mathop  = \limits^{(a)}  - {\text{Tr}}({({\sigma ^2}{{\mathbf{I}}_{{N_{{\text{rx}}}}}} + {{{\mathbf{\tilde H}}}_k}{\mathbf{\tilde F}}{{\mathbf{S}}_k}{\mathbf{S}}_k^H{{{\mathbf{\tilde F}}}^H}{\mathbf{\tilde H}}_k^H)^{ - 1}} \times  \hfill  \\
  {\rm d}({\sigma ^2}{{\mathbf{I}}_{{N_{{\text{rx}}}}}} + {{{\mathbf{\tilde H}}}_k}{\mathbf{\tilde F}}{{\mathbf{S}}_k}{\mathbf{S}}_k^H{{{\mathbf{\tilde F}}}^H}{\mathbf{\tilde H}}_k^H))  \\
  \mathop  = \limits^{(b)}  - {\text{Tr}}({({\sigma ^2}{{\mathbf{I}}_{{N_{{\text{rx}}}}}} + {{{\mathbf{\tilde H}}}_k}{\mathbf{\tilde F}}{{\mathbf{S}}_k}{\mathbf{S}}_k^H{{{\mathbf{\tilde F}}}^H}{\mathbf{\tilde H}}_k^H)^{ - 1}}{{{\mathbf{\tilde H}}}_k}{\mathbf{\tilde F}}{{\mathbf{S}}_k}{\mathbf{S}}_k^H{\rm d}({{{\mathbf{\tilde F}}}^H}){\mathbf{\tilde H}}_k^H) \hfill \\
  \mathop  = \limits^{(c)}  - {\text{Tr}}({\mathbf{\tilde H}}_k^H{({\sigma ^2}{{\mathbf{I}}_{{N_{{\text{rx}}}}}} + {{{\mathbf{\tilde H}}}_k}{\mathbf{\tilde F}}{{\mathbf{S}}_k}{\mathbf{S}}_k^H{{{\mathbf{\tilde F}}}^H}{\mathbf{\tilde H}}_k^H)^{ - 1}}{{{\mathbf{\tilde H}}}_k}{\mathbf{\tilde F}}{{\mathbf{S}}_k}{\mathbf{S}}_k^H{\rm d}({{{\mathbf{\tilde F}}}^H})), \hfill \\ 
\end{gathered} 
\end{equation}
where $(a)$ is based on ${\rm d}(\log \det ({\mathbf{A}})) = {\text{Tr(}}{{\mathbf{A}}^{ - 1}}{\rm d}({\mathbf{A}}){\text{)}}$, $(b)$ is obtained since the differential should be computed with respect to ${{{\mathbf{\tilde F}}}^ C }$ with $\mathbf{\tilde F}$ treated as a constant matrix, and $(c)$ is due to $\text{Tr}(\mathbf{A}\mathbf{B})=\text{Tr}(\mathbf{B}\mathbf{A})$ again.

Substituting (\ref{eqa7}) into (\ref{eqa6}), the lemma readily follows.   $\hfill\blacksquare$

Based on the result in \textbf{Lemma 4}, we can then derive the conjugate gradient of the objective function in (\ref{eqa1}a), i.e., the weighted-sum of user rates, with respect to $\mathbf{\tilde F}$, as given by
\begin{equation} \label{eqa9}
\begin{gathered}
  \sum\limits_{k = 1}^K {{\omega _k}} \nabla_{\mathbf{\tilde F}} {f_k}({\mathbf{\tilde F}}) 
  =  - \sum\limits_{k = 1}^K \omega _k\mathbf{\tilde H}_k^H\sigma^2\mathbf{I}_{N_{\text{rx}}} +\hfill \\
    \quad\quad\quad\quad\quad\quad \mathbf{\tilde H}_k\mathbf{\tilde F}\mathbf{S}_k
    \mathbf{S}_k^H\mathbf{\tilde F}^H\mathbf{\tilde H}_k^H)^{ - 1}
    \mathbf{\tilde H}_k\mathbf{\tilde F}\mathbf{S}_k\mathbf{S}_k^H  \hfill \\ 
\end{gathered} 
\end{equation}

Based on the conjugate gradient expression (\ref{eqa9}), a Riemannian gradient descent algorithm can be run with the existing \emph{Manopt} toolbox to find a stationary point solution $\mathbf{\tilde{F}}^{*}$ for (\ref{eqa1}). Note that when there is only $K=1$ user, i.e., in the single-user case, the conjugate gradient in (\ref{eqa9}) reduces to a very simple expression: $ - {{\mathbf{\tilde H}}^H}{({\sigma ^2}{{\mathbf{I}}_{{N_{{\text{rx}}}}}} + {\mathbf{\tilde H}}{{\mathbf{\tilde F}}_{\text{r}}}{\mathbf{\tilde F}}_{\text{r}}^H{{\mathbf{\tilde H}}^H})^{ - 1}}{\mathbf{\tilde H}}{{\mathbf{\tilde F}}_{\text{r}}}$, based on which the Riemannian gradient descent algorithm can be also run to approximately solve (\ref{eq7}).

With $\mathbf{\tilde{F}}^{*}$, we can in turn obtain a stationary point solution ${{\mathbf{F}}^*} = {\mathbf{L}}{{\mathbf{\tilde F}}^*}$ for problem (\ref{eq9}). 

\section{Proof of Lemma 2}
For problem (\ref{eq10}) with any fixed $\mathbf{F}$, it is clear the optimal $\mathbf{G}$ is provided by the MMSE one in (\ref{eq11}) and the optimal $\mathbf{W}$ is given by (\ref{eq12}).  Substituting (\ref{eq11}) and (\ref{eq12}) in (\ref{eq10}), the problem can be then reduced into
\begin{subequations}  \label{eqb1}
\begin{align}
  \mathop {\min }\limits_{\mathbf{F}} &\sum\limits_{k = 1}^K {{\omega _k}\log \det ({\mathbf{E}}_k^{{\text{mmse}}})}  \\
  {\text{s}}{\text{.t}}{\text{.   }}&{\mathbf{F}}{{\mathbf{F}}^H} = {{\mathbf{R}}_{{\text{des}}}},
\end{align} 
\end{subequations}
where ${{\mathbf{E}}_k^{{\text{mmse}}}}$ is defined in (\ref{eq11a}). 
By examining the term $\log \det ({{\mathbf{E}}_k^{{\text{mmse}}}})$ in (\ref{eqb1}a), we have 
\begin{equation}  \label{eqb2}
\begin{aligned}
  &\log \det ({{\mathbf{E}}_k^{{\text{mmse}}}})=-\log \det(({{\mathbf{E}}_k^{{\text{mmse}}}})^{-1})  \\
  \mathop  = \limits^{(a)} &-\log \det ({({{\mathbf{I}}_d} - {\mathbf{F}}_k^H{\mathbf{H}}_k^H{({{\mathbf{H}}_k}{{\mathbf{R}}_{{\text{des}}}}{\mathbf{H}}_k^H + {\sigma ^2}{{\mathbf{I}}_{{N_{{\text{rx}}}}}})^{ - 1}}{{\mathbf{H}}_k}{{\mathbf{F}}_k})^{ - 1}})  \\
   \mathop  = \limits^{(b)} &-\log \det ({{\mathbf{I}}_d} + {\mathbf{F}}_k^H{\mathbf{H}}_k^H{({\sigma ^2}{{\mathbf{I}}_{{N_{{\text{rx}}}}}}{\text{ + }}\sum\limits_{i \ne k}^{} {{{\mathbf{H}}_k}{{\mathbf{F}}_i}{\mathbf{F}}_i^H{\mathbf{H}}_k^H}+} \\ 
   &{{{\mathbf{H}}_k}{{\mathbf{F}}_{\text{r}}}{\mathbf{F}}_{\text{r}}^H{\mathbf{H}}_k^H)^{ - 1}}{{\mathbf{H}}_k}{{\mathbf{F}}_k})=-{C_k}
\end{aligned} 
\end{equation}
where $(a)$ follows by applying (\ref{eq11a}) and $(b)$ holds due to the Woodbury matrix identity. It is shown in (\ref{eqb2}) that $\log \det ({{\mathbf{E}}_k^{{\text{mmse}}}})$ is actually the negative of the maximum achievable rate $C_k$ of user $k$ in (\ref{eq3a}). It then readily follows that problem (\ref{eq10}) is equivalent to problem (\ref{eq9}) in the sense that the optimal solution ${{\mathbf{F}}^*}$ of two problems are identical.

We proceed to show ${{\mathbf{F}}^*}$ is a stationary point solution for (\ref{eq9}) if and only if it is part of a stationary point solution $(\mathbf{F}^*, \mathbf{G}^*, \mathbf{W}^*)$ for (\ref{eq10}). For brevity, we re-define (\ref{eq10}a) as ${g_1}({\mathbf{F}},{\mathbf{G}},{\mathbf{W}})$ and (\ref{eqb1}a) as $g_2({\mathbf{F}})$. Assuming $({{\mathbf{F}}^*},{{\mathbf{G}}^*},{{\mathbf{W}}^*})$ is  a stationary point for (\ref{eq10}), we immediately have
\begin{equation} \label{eqb4}
    {\text{Tr}}({\nabla _{\mathbf{F}}}{g_1}{({{\mathbf{F}}^*},{{\mathbf{G}}^*},{{\mathbf{W}}^*})^H}({\mathbf{F}} - {{\mathbf{F}}^*})) \leqslant 0,
\end{equation}
where $\mathbf{F}$ satisfies ${\mathbf{F}}{{\mathbf{F}}^H} = {{\mathbf{R}}_{{\text{des}}}}$.

Using basic differentiation rule, we derive the differential  of $g_1$ with respect to $\mathbf{F}^C$ as
\begin{equation}  \label{eqb3}
\begin{aligned}
  {\rm{d}}({g_1}({{\mathbf{F}}^*},{{\mathbf{G}}^*},{{\mathbf{W}}^*}))\mathop  = \limits^{(a)} &\sum\limits_{k = 1}^K {{\omega _k}{\text{(Tr}}({\mathbf{W}}_k^*{\rm{d}}({{\mathbf{E}}_k}))} )  \\
  \mathop  = \limits^{(b)} &\sum\limits_{k = 1}^K {{\omega _k}{\text{(Tr}}({{({\mathbf{E}}_k^{{\text{mmse}}})}^{ - 1}}{\rm{d}}({\mathbf{E}}_k^{{\text{mmse}}}))} ) \\
  \mathop  = \limits^{(c)}& {\rm{d}}({g_2}({{\mathbf{F}}^*})),
\end{aligned} 
\end{equation}
where $(a)$ and $(c)$ hold based on the chain rule and $(b)$ is obtained by applying (\ref{eq12}). From (\ref{eqb3}), we have  
\begin{equation}  \label{eqb5}
{\nabla _{\mathbf{F}}}{g_1}({{\mathbf{F}}^*},{{\mathbf{G}}^*},{{\mathbf{W}}^*}) = {\nabla _{\mathbf{F}}}{g_2}({{\mathbf{F}}^*}).
\end{equation}
Combining (\ref{eqb4}) and (\ref{eqb5}), we can derive
\begin{equation} 
    {\text{Tr}}({\nabla _{\mathbf{F}}}{g_2}{({{\mathbf{F}}^*})^H}({\mathbf{F}} - {{\mathbf{F}}^*})) \leqslant 0,
\end{equation}
implying $\mathbf{F}^*$ is a stationary point solution for problem (\ref{eqb1}), as well as its equivalent  problem (\ref{eq9}).
 
Likewise, we can prove
that $({{\mathbf{F}}^*},{{\mathbf{G}}^*},{{\mathbf{W}}^*})$ generating from a stationary point solution $\mathbf{F}^*$ for problem (\ref{eq9}) is also a stationary point solution for problem~(\ref{eqb1}).
Then, Lemma 2 can be approved. 

\section{Comparison with the SDP Approach~\cite{12}}
Here, we show the SDP method proposed in \cite{12} with additional $D = d \times K$ radar beamformers does not bring any benefit in achievable weighted sum-rate over our closed-form solution for (\ref{eq13}). 
This starts with the following lemma:

\emph{\textbf{Lemma 5}\cite[Theorem 1]{12}: Assume a communication beamforming matrix ${{\mathbf{P}}_{\rm c}} = [{{\mathbf{P}}_1},{{\mathbf{P}}_2},\cdots,{{\mathbf{P}}_K}] \in {\mathbb{C}^{{N_{\rm tx}} \times D}}$ and a radar  beamforming matrix ${{\mathbf{P}}_{\rm r}} \in {\mathbb{C}^{N_{\rm tx} \times N_{\rm tx}}}$. Given the transmit covariance ${{\mathbf{R}}_{\rm{des}}}$, there exist ${{\mathbf{P}}_{\rm c}}$ and ${{\mathbf{P}}_{\rm r}}$ that satisfy
\begin{equation}    \label{eq18}
{{\mathbf{R}}_{\rm{des}}}={{\mathbf{P}}_{\rm c}}{\mathbf{P}}_{\rm c}^H + {{\mathbf{P}}_{\rm r}}{\mathbf{P}}_{\rm r}^H,
\end{equation}
if and only if 
\begin{equation}  \label{eq19}
{{\mathbf{P}}_{\rm c}}{\mathbf{P}}_{\rm c}^H \preceq {{\mathbf{R}}_{{\rm{des}}}}.
\end{equation}
Given such a ${{\mathbf{P}}_{\rm c}}$, we can construct ${{\mathbf{P}}_{\rm r}}$ by 
\begin{equation}  
{{\mathbf{P}}_{\rm r}}= {({{\mathbf{R}}_{{\rm{des}}}} - {{\mathbf{P}}_{\rm c}}{{{\mathbf{P}}_{\rm c}}^H})^{1/2}}.
\end{equation}
}

By \textbf{Lemma 5}, the (non-convex) transmit covariance constraint (\ref{eq18}) can be equivalently transformed into a convex semi-definite constraint (\ref{eq19}). Let $\mathbf{P}=[\mathbf{P}_{\rm c}, \mathbf{P}_{\rm r}]$. With $\mathbf{F}$ replaced by $\mathbf{P}$, 
problem (\ref{eq13}) can be recast as 
\begin{subequations} \label{eq20}
\begin{align}
  \mathop {\max }\limits_{{{\mathbf{P}}}_{\rm c}} {\text{ }}& \sum\limits_{k = 1}^K {{\omega_k}{\text{Re\{Tr}}({{\mathbf{W}}_k}{\mathbf{G}}_k^H{{\mathbf{H}}_k}{{\mathbf{P}}_k})} \}   \\
  {\text{s}}{\text{.t}}{\text{.   }}& {{\mathbf{P}}_{\rm c}}{\mathbf{P}}_{\rm c}^H \preceq {{\mathbf{R}}_{{\rm{des}}}}.
\end{align}
\end{subequations}
Again, defining  ${\mathbf{\tilde P}} = {{\mathbf{L}}^{ - 1}}{\mathbf{P}}$ and ${\mathbf{\tilde H}_k} = {\mathbf{H}_k \mathbf{L}}$, we can   equivalently reformulate (\ref{eq20}) as 
\begin{subequations}  \label{eq21}
\begin{align}
  \mathop {\max }\limits_{{\mathbf{\tilde P}}_{\rm c}} {\text{ }}& {\text{ Re\{Tr}}({\mathbf{M}^H}{{{\mathbf{\tilde P}}}_{\rm c}})\}   \\
  {\text{s}}{\text{.t}}{\text{.   }}& {{{\mathbf{\tilde P}}}_{\rm c}}{{{\mathbf{\tilde P}}}_{\rm c}}^H \preceq {{\mathbf{I}}_{N_{\rm tx}}},
\end{align}
\end{subequations}
which is a convex SDP. Existing convex solvers can then be employed to efficiently compute the globally optimal ${\mathbf{\tilde P}}_{\rm c}^ * $ for (\ref{eq21}). Interestingly, we can show:

\emph{Proposition 2: The optimal ${\mathbf{\tilde F}}_{\rm c}^ * $ for (\ref{eq14}) is identical to the optimal ${\mathbf{\tilde P}}_{\rm c}^ * $ for (\ref{eq21}). }

\emph{Proof:} By Schur Complement, problem (\ref{eq21}) can be equivalently rewritten as  
\begin{subequations} \label{eq22}
\begin{align}
  \mathop {\max }\limits_{{{{\mathbf{\tilde P}}}_{\rm c}}} &{\text{ Re\{Tr}}({\mathbf{M}^H}{{{\mathbf{\tilde P}}}_{\rm c}})\}  \\
  {\text{s}}{\text{.t}}{\text{. }} &{{{\mathbf{\tilde P}}}_{\rm c}}^H{{{\mathbf{\tilde P}}}_{\rm c}} \preceq {{\mathbf{I}}_D} ,
\end{align}
\end{subequations}
Compared to problem (\ref{eq14}), this problem has an identical objective and slightly different constraint ${{{\mathbf{\tilde P}}}_{\rm c}}^H{{{\mathbf{\tilde P}}}_{\rm c}} \preceq {{\mathbf{I}}_D}$. Next, we rely on the Karush-Kuhn-Tucker (KKT) optimality condition~\cite{17} to show that  the optimal ${\mathbf{\tilde P}}_{\rm c}^*$ for (\ref{eq22}) always satisfies ${({\mathbf{\tilde P}}_{\rm c}^*)^H}{\mathbf{\tilde P}}_{\rm c}^* = {{\mathbf{I}}_D}$.

For the convex problem (\ref{eq22}), the Lagrange function is 
\begin{equation}  
\mathcal{L}({{\mathbf{\tilde P}}_{\rm c}},{\mathbf{Q}}) = {\text{Re\{Tr}}({\mathbf{M}^H}{{\mathbf{\tilde P}}_{\rm c}})\}  + {\text{Tr}}({\mathbf{Q}}({{\mathbf{I}}_D} - {{\mathbf{\tilde P}}_{\rm c}}^H{{\mathbf{\tilde P}}_{\rm c}})),
\end{equation}
where the dual variable ${\mathbf{Q}} \in {\mathbb{C}^{D \times D}}$ is a positive semi-definite matrix, i.e.,  ${\mathbf{Q}} \succeq {\mathbf{0}}$. Let $\mathbf{Q}^*$ denote the optimal dual matrix. By the KKT condition, we must have
\begin{equation} 
\mathbf{M}^H- \mathbf{Q}^* ({{\mathbf{\tilde P}}_{\rm c}^*})^H =\mathbf{0}.
\end{equation}
This implies that each column of ${\mathbf{M}^H}$ should lie in the range of ${\mathbf{Q}^*}$, i.e. ${\mathbf{M}^H} \in \cal{R}({\mathbf{Q}^*})$. In other words, ${\rm rank}({\mathbf{M}}) \leqslant {\rm rank}({\mathbf{Q}^*})$. Recall that $\mathbf{M}$ is full rank, i.e.,  ${\rm rank}({\mathbf{M}}) = D$ in almost all cases. This in turn infers  ${\rm rank}({\mathbf{Q}^*}) = D$, or $\mathbf{Q}^*  \succ \mathbf{0}$. By complementary slackness condition, we also have
\begin{equation} 
{\text{Tr}}({{\mathbf{Q}}^*}({{\mathbf{I}}_D} - {({\mathbf{\tilde P}}_{\rm c}^*)^H}{\mathbf{\tilde P}}_{\rm c}^*) = 0.
\end{equation}
As $\mathbf{Q}^* \succ \mathbf{0}$, we must have
\begin{equation}  
{({\mathbf{\tilde P}}_{\rm c}^*)^H}{\mathbf{\tilde P}}_{\rm c}^* = {{\mathbf{I}}_D}.
\end{equation}

It then follows that ${\mathbf{\tilde P}}_{\rm c}^*$ actually satisfies the constraint (\ref{eq14}b). This simply means that $ {\mathbf{\tilde P}}_{\rm c}^*$ is also an optimal solution to problem (\ref{eq14}). Due to the uniqueness of the optimal solution for (\ref{eq14}), it readily follows ${\mathbf{\tilde P}}_{\rm c}^*={\mathbf{\tilde F}}_{\rm c}^*$.  $\hfill\blacksquare$
}

Given ${\mathbf{\tilde P}}_{\rm c}^*={\mathbf{\tilde F}}_{\rm c}^*$, we have ${\mathbf{P}}_{\rm c}^*={\mathbf{F}}_{\rm c}^*$. Note that the radar beamformers in $\mathbf{F}_{\rm r}^*$ and $\mathbf{P}_{\rm r}^*$ are only useful in satisfying the transmit covariance constraint to guarantee radar performance, but have no impact on communication rates. In addition, it can be seen that the updates of the optimal MMSE receiver $\mathbf{G}^{\rm mmse}$ in (\ref{eq11}) and the optimal weight matrix $\mathbf{W}^*$ in (\ref{eq12}) only depend on the given $\mathbf{F}_{\rm c}$. Therefore, Algorithm 1, using the SDP approach [12], would converge to the same communication beamforming matrix $\mathbf{F}_{\rm c}^*$, as the proposed one. Hence, we conclude that the SDP approach proposed in \cite{12} with extra computational and implementation complexity cannot bring any benefit in achievable weighted sum-rate.


%

\bibliographystyle{IEEEtran}
\bibliography{citex}

\newpage


\vfill

\end{document}